\documentclass[review,3p]{elsarticle}

\usepackage{lineno,hyperref}
\usepackage{amsmath}
\usepackage{amsfonts}
\usepackage[english]{babel}
\usepackage[utf8]{inputenc}
\usepackage{color}
\usepackage{graphicx}
\usepackage[version=4]{mhchem}
\usepackage{rotating}

\usepackage{fullpage}
\usepackage{mathtools}
\usepackage{caption}
\usepackage{tikz}
\usepackage{standalone}
\usetikzlibrary{decorations.pathreplacing}
\usetikzlibrary{arrows}
\usetikzlibrary{fadings}
\usepackage{amssymb}
\usepackage{amsmath}
\modulolinenumbers[5]
\usepackage{xcolor}

\newcommand{\pd}[2]{\frac{\partial #1}{\partial #2}}
\newcommand{\fd}[2]{\frac{\text{d} #1}{\text{d} #2}}
\newcommand{\bs}[1]{\boldsymbol #1}
\newcommand{\kd}[1]{\delta_{#1}} 
\newcommand{\del}{\bs{\nabla}}
\newcommand{\tr}{\text{tr}}

\newcommand{\C}{\text{C}}
\newcommand{\rad}{R}
\newcommand{\volfrac}{\psi}
\newcommand{\maxconc}{\check{c}_0}

\newcommand{\expvol}{V}

\journal{Journal of The Electrochemical Society}

\begin{document}

\begin{frontmatter}

\title{The effect of mechanical stress on lithium distribution and geometry optimisation for multi-material lithium-ion anodes.}

\author{Ian P. E. Roper$^{\text{a},*}$, S. Jon Chapman$^\text{b,c}$, Colin P. Please$^\text{b,c}$}
\address{$^\text{a}$ InFoMM CDT, Mathematical Institute, Woodstock Road, Oxford, UK\\
$^\text{b}$ OCIAM, Mathematical Institute, Woodstock Road, Oxford, UK\\
$^\text{c}$ Faraday Institution}
{\let\thefootnote\relax\footnote{{$^*$ Corresponding author\\ \textit{Email addresses:} \texttt{i.roper1993@gmail.com} (Ian P. E. Roper)\\ \texttt{chapman@maths.ox.ac.uk} (S. Jon Chapman)\\ \texttt{please@maths.ox.ac.uk} (Colin P. Please)}}}




\begin{abstract}
 A model is presented for predicting the open-circuit voltage (OCV) and lithium distribution within lithium-ion anodes containing multiple materials, coupling linear elasticity with a stress-dependent chemical potential.
 The model is applied to a spherical radially-symmetric nano-particle with a silicon core and a graphite shell, highlighting the large effect on lithium distribution and OCV caused by the stress-coupling.
 Various performance measures based on the expanded volume, the amount of lithium intercalated and the maximum stress induced, are calculated for a silicon core with a graphite shell to enable optimisation of the volume of the silicon core.
\end{abstract}


\end{frontmatter}


 \section{Introduction}
 
 Lithium-ion batteries are ubiquitous in modern technology, powering mobile phones, laptop computers and electric vehicles \cite{Whittingham2012}.
 Many different materials are being tested for use in all components of lithium-ion batteries to achieve the largest energy capacity while remaining portable and light-weight.
 A key component for maximising energy capacity is the anode; the more lithium that can be intercalated into the anode, the more charge can be stored, and thus the more energy can be obtained from a single charge.
 However, upon lithiation many high capacity anodes increase in volume \cite{zhang2011review}, due to changes in the crystal structure.
 This can cause mechanical stresses because of geometrical constraints within the anode and the rest of the battery, but also because of concentration gradients of lithium within the anode causing non-uniform expansion.
 The expansion of the anode can displace other components of the battery and the high stresses within the anode material itself can cause it to crack \cite{beaulieu2001colossal}.
 These effects can result in loss of connection between components within the battery, leading to capacity fade after several charge/discharge cycles.
 
 One anode material with high expansion and a large capacity for lithium atoms which has received a lot of recent research attention is silicon.
 When fully lithiated, silicon can accommodate 3.75 lithium atoms per silicon atom, forming the alloy Li$_{3.75}$Si at room temperature \cite{Hatchard2004}, giving silicon anodes a volumetric capacity of 3500 mAh/g.
 This is much greater than the volumetric capacity of 372 mAh/g achieved by commonly-used graphite anodes \cite{liu2016well}, due to graphite only being able to accommodate one lithium atom per six carbon atoms (LiC$_6$).
 This makes silicon a very inviting anode material; however, when silicon is lithiated, it expands to around four times its original size \cite{chakraborty2015combining}, much more than the $\sim10$\% volume increase observed in fully lithiated graphite.
 
 Experimental efforts to make silicon a viable anode material often involve using nano-structures within the anode design to minimise the variation in the lithium concentration.
 These designs are often based on nanoparticles \cite{kim2010critical,liu2012size, McDowell2013,song2014micro}, nanowires \cite{chan2008high,hieu2012free,Zamfir2013a} and nanotubes \cite{Song2010a,wang2016tuning}.
 Additionally, there are many nano-structure designs which attempts to constrain the expansion of the silicon, for example in a core--shell structure \cite{song2014micro,Xu2010,Yoshio2002}, or yolk--shell structure \cite{Liu2012b,Wang2016b}, or using self-healing polymers \cite{Wang2013a}.
 These structures are often used in conjunction with each other, for example Wang \textit{et al.} showed that for a silicon oxide shell which is thick enough, inwards growth can be induced in silicon nanotubes \cite{wang2016tuning}.
 
 In determining which structures may be the most effective at mitigating the adverse effects of highly expanding anode materials, it is insightful to construct mathematical models for the mechanical behaviour of the anode and the behaviour of the lithium within.
 Several mathematical models for the expansion of anodes upon lithiation have been proposed.
 In general, the expansion is assumed to be due to the intercalation of lithium, and opposed by a mechanical response from the stiffness of the anode material, causing stresses.
 Linear elasticity models \cite{Chen2016,Ryu2011a,zhang2007numerical} have been used to model the mechanical response when strains are small, but more commonly finite strain models using geometrically nonlinear elasticity are used for silicon to capture the large strains that occur \cite{Bower2010a,chakraborty2015combining,Chen2014,cui2012finite,Gao2013a}.
 Sethuraman \textit{et al.} also showed that silicon plastically flows during cycling \cite{VijayA.SethuramanMichaelJ.Chon2010a} due to the high stresses induced. 
 This plastic flow has been incorporated into several finite strain models \cite{Bower2010a,chakraborty2015combining,Chen2014}, as well as linear models \cite{Wu2018}.
 Several more simple models have neglected the elastic stresses altogether and modelled the mechanical response of the silicon as solely plastic \cite{Jia2016,Zhao2012}.
 An additional phenomenon that has been the basis of several modelling studies is crack formation and propagation due to the high stresses within the anode \cite{Chen2016,Esmizadeh2019,Gao2013a,Sarkar2017,Wu2018}.
 
 Two types of model for the lithiation are commonly used: a single phase model in which the silicon is gradually lithiated \cite{Bower2010a,chakraborty2015combining,cui2012finite}, and two-phase models in which there is a sharp reaction front between the lithiated phase and the unlithiated phase \cite{Huang2013,Wu2018,Zhang2018}.
 The former can be simply modelled using diffusion of the lithium atoms, whereas the lithium kinetics in the latter are often described using a Cahn-Hilliard phase-field model.
 Experimentally, the two-phase model has been shown to likely be more physically accurate \cite{Chen2014,McDowell2013}.
 The time-dependence of these models for the evolution of the lithium concentration often cause the main focus of these studies to be on the stresses inside the nano-structures due to silicon non-uniformity, rather than due to material heterogeneity \cite{Chen2016,Chen2014,cui2012finite,Ryu2011a}.
 
 The coupling of the lithium distribution to the stress is a two-way phenomenon, since in addition to lithiation causing expansion, lithium diffusion is also affected by stress.
 Stress-assisted diffusion is typically included into lithium transport models by considering the chemical potential of the lithium in the anode to depend on the hydrostatic stress.
 This coupling affects the diffusion of the lithium through the anode due to the stress induced by lithiation.
 Several works \cite{Bower2010a,Chen2016,Cui2012b,gao2011strong,Ryu2011a} have included stress-assisted diffusion using the model formulated by Larche and Cahn \cite{Larche1973a} using a small-strain assumption.
 Works by Wu \cite{wu2001a} and Cui \textit{et al.} \cite{cui2012finite} have resulted in a more general framework for incorporating a stress-dependence into the chemical potential of the lithium.
 However, Cui's general form of stress-dependent chemical potential reduces to those proposed by Wu, Larche and Cahn under the appropriate assumptions.
 This general stress-assisted diffusion framework has more recently been applied to finite strain models, for example in \cite{chakraborty2015combining}.
 
 To simplify the mathematical models of these anode nano-structures, simplified geometries are often used.
 Nano-particles are approximated by spherical geometries \cite{cui2012finite,Wu2018,Zhao2012}, nano-wires and nano-rods are approximated to cylindrical geometries \cite{chakraborty2015combining}, and thin films of silicon are assumed to grow unidirectionally \cite{bucci2014measurement}.
 However, even in these simplified geometries, the use of nonlinear elasticity or other complex models requires the use of software packages to solve the resulting differential equations \cite{chakraborty2015combining,Chen2014,cui2012finite}.
 Despite these systems requiring numerical solvers, the chemical potential of the lithium is often approximated as an analytical expression of the lithium concentration \cite{chakraborty2015combining,cui2012finite,sethuraman2010situ}.
 These approximations fail to capture the large `steps' in chemical potential that occur at transitions in the structure of the intercalated anode \cite{Birkl2015a}.
 Using an interpolated function of the chemical potential of the lithium in a material would capture these steps and their use in these complicated models would only marginally increase the computational cost.
 
 In this work, we consider a multi-material anode, which we call a hybrid anode.
 We present a linearly elastic mechanical model coupled to a stress-assisted diffusion model using the stress-dependent chemical potential proposed by Larche and Cahn with a stress-independent term depending on lithium concentration.
 We focus on a time-scale much slower than the diffusion of lithium through the nano-particle allowing us to use a quasi-static approximation, focusing on the stresses induced by the different expansions of the different materials in chemical and mechanical equilibrium.
 We present this model as a general chemo-mechanical model for anode lithiation.
 We then apply our model to a spherical, radially-symmetric geometry consisting of a core and shell of different materials, representing a nano-particle with a constraining shell.
 
 Previous works have often modelled a constrained core--shell anode nano-particle using a zero displacement condition at the shell interface \cite{cui2012finite,Zhao2012} or by modelling the shell as a material that cannot be lithiated \cite{Hao2013,Wang2016a}.
 One of our main focuses is the distribution of lithium between the core and the shell and thus we include the lithiation of both materials and the resulting displacement of the interface.
 We show how the distribution of lithium between the two materials depends heavily on both stress-assisted diffusion and the geometry of the nano-particle.
 
 The lithium distribution predicted by the model is used to calculate the chemical potential of the lithium in the hybrid nano-particle, which we can then convert to an OCV.
 OCVs can be used to infer the state of charge (SOC) of a battery at a given voltage and are used in battery management systems and other battery models \cite{Dubarry2012,Wang2014}.
 While experimentally measured OCVs for single material anodes are easily found within the literature \cite{Reynier2003a,li2007situ}, experimentally obtaining the OCV curve of a new anode design with multiple materials can be expensive and is very time-consuming.
 Therefore, a theoretical technique for predicting OCVs of multi-material anodes is of interest to manufacturers and researchers.
 Sethuraman showed that stresses can significantly affect the OCVs \cite{sethuraman2010situ}, thus the OCV of a hybrid anode is not only dependent on the individual materials but also on the stresses induced in each material by the others at equilibrium.
 We show the importance of including stress into the lithium concentration model by comparing chemical potentials when including and excluding stress-assisted diffusion.
 
 Lastly, while the modelling of the stresses within lithium-ion anodes has been widely studied, very few of these models discuss optimal designs based on the results of the models.
 In the second half of this work, we discuss several different performance measures that could be used in defining optimality for the anode design.
 These are i) the amount of lithium intercalated, ii) the expanded volume of the anode, and iii) the maximum induced stress.
 We then use these performance measures to suggest three different optimality conditions for the hybrid anode design.
 We derive these conditions for the silicon core--graphite shell nano-particle design and show how the objective function varies with the volume of the silicon core and find the optimal design in each of the three cases.
 
 The rest of this paper is organised as follows.
 In Section~\ref{subsec:General_formulation} we formulate the chemo-mechanical model we use to describe the lithiation of a lithium-ion anode for a general multi-material anode.
 In Section~\ref{subsec:using_OCVs}, we describe how the stress-free chemical potential of the lithium in each anode material can be calculated from the material's OCV.
 We then simplify the general model to one for a radially symmetric nano-particle with a core of material 1 and a shell of material 2 and solve this model, showing that for certain simple geometries, the concentrations are uniform in each material, even with the inclusion of stress-assisted diffusion.
 In Section~\ref{sec:Chemical_results}, we present the lithium concentrations and chemical potentials in a nano-particle with a silicon core and a graphite shell to show the importance of including stress-assisted diffusion into the lithium concentration model.
 In Section~\ref{subsec:Performance_measures} we then discuss the three performance measures we will use to then find an optimal design of the silicon--graphite nano-particles in Section~\ref{subsec:optimisation}.
 Finally in Section~\ref{sec:conclusions}, we discuss the conclusions that can be drawn from this work and the validity of the results in certain regimes.
 
 \section{Mathematical Model}\label{sec:model}
 
 In this section, we develop a mathematical model for the lithiation of a multi-material anode.
 We begin with a general geometry and arbitrary number of different anode materials.
 We then apply the model to a spherical, radially symmetric geometry with a core of one material and a shell of another material.
 The anode is the only battery component which we model in this work; we do not model the electrolyte, the cathode, binder, current collectors or any other components.
 
 \subsection{General Formulation}\label{subsec:General_formulation}
 
 We denote the domain of the entire anode as $\Omega$ and the region occupied by each anode material as $\Omega_a$ with $a = 1, \dots, n$, where $n$ is the number of different materials in the hybrid anode.
 We present boundary conditions on the boundary between the anode and the electrolyte $\Gamma_e$ and on boundaries between different anode materials $\Gamma_{ab}$ where $a \neq b$ and $a, b \in \{1, \dots, n\}$.
 
 \subsubsection{Mechanical Model}\label{subsubsec:Mechanical_model}
 
 We suppose the deformation in the anode is due to the volumetric changes due to lithiation and is counteracted by an elastic response from the material due to its stiffness.
 For simplicity, we assume the strains are small enough that we may use linear elasticity theory.
 Similar to previous linearly-elastic studies of anode expansion \cite{Chen2016,zhang2007numerical}, we write the elastic strain tensor as
 \begin{equation}
 \textbf{E}^e = \frac{1}{2}\Big[\big(\del^*\bs{u}^*\big)^\text{T} + \del^*\bs{u}^*\Big] - \eta_aV^m_ac^*_a\bs{1} \quad \text{ in } \Omega_a, \label{eq:linear_elastic_strain_tensor}
 \end{equation}
 where $\bs{u}^* = (u_1^*, u_2^*, u_3^*)$ is the displacement (m), $c_a^*$ is the concentration of lithium (mol m$^{-3}$) and $\del^*$ is the gradient operator.
 The molar volume of material $a$, measured when zero lithium has been intercalated, is denoted as $V^m_a$ (m$^3$ mol$^{-1}$) and $\eta_a$ is the coefficient of compositional expansion (CCE) of material $a$, which is a measure of the volumetric expansion due to lithiation \cite{swaminathan2007electrochemomechanical}.
 
 We relate the stress to the strain tensor by Hooke's law,
 \begin{equation}
 \bs{\sigma}^* = \mathbb{C}_a^*(c_a^*) : \textbf{E}^e \quad \text{ in } \Omega_a,\label{eq:stress-strain_relation} 
 \end{equation}
 where $\mathbb{C}_a^*$ is the stiffness tensor of material $a$ which can vary with the lithiation state of the material.
 We assume that each anode material is isotropic, allowing us to write the stiffness tensor in suffix notation as
 \begin{equation}
 \mathbb{C}^*_a(c_a^*) = {c_{ijkl}^{i,*}}(c_a^*) = \lambda_a^*(c_a^*)\kd{ij}\kd{kl} + G_a^*(c_a^*)\kd{ik}\kd{jl} + G_a^*(c_a^*)\kd{il}\kd{jk},
 \end{equation}
 where $\lambda^*_a$ is the first Lam\'{e} parameter of material $a$ and $G_a^*$ is the shear modulus of material $a$ \cite{howell2009applied}, both of which can vary with the lithiation state of the material.
 We now write \eqref{eq:stress-strain_relation} as
 \begin{equation}
 \bs{\sigma}^* = \mathbb{C}^*_a(c_a^*) : \big(\del^*\bs{u}^* - \eta_aV^m_ac^*_a\bs{1}\big) \quad \text{ in } \Omega_a,\label{eq:dimensional_stress}
 \end{equation}
 which in suffix notation, becomes
 \begin{equation}
 \sigma^*_{ij} = \lambda_a^*(c_a^*)\kd{ij}\pd{u_k^*}{x_k^*} + G_a^*(c_a^*)\bigg(\pd{u_i^*}{x_j^*} + \pd{u_j^*}{x_i^*}\bigg) - \eta_aV^m_ac^*_a(3\lambda_a^*(c_a^*)+2G_a^*(c_a^*))\kd{ij} \quad \text{ in } \Omega_a, \label{eq:dimensional_stress_suffix}
 \end{equation}
 where we use the summation convention for repeated indices.
 Lastly, the entire anode is in mechanical equilibrium, which is given by
 \begin{equation}
 \del^* \cdot \bs{\sigma}^* = 0 \quad\quad\quad \text{ in }\Omega.\label{eq:dimensional_mechanical_equilibrium}
 \end{equation}
 We prescribe a traction free boundary condition on $\Gamma_e$, giving
 \begin{equation}
 \bs{\sigma}^* \cdot \textbf{n}_e = 0 \quad\quad\quad \text{ on }\Gamma_e, \label{eq:dimensional_traction_free}
 \end{equation}
 where $\textbf{n}_e$ is the unit vector normal to the interface between the anode and the electrolyte.
 We also prescribe continuity of normal stress and displacement between anode materials
 \begin{equation}
 [\bs{\sigma}^* \cdot \textbf{n}_{ab}]^+_- = [\bs{u}^*]^+_- = 0 \quad\quad\quad \text{ on }\Gamma_{ab}, \label{eq:dimensional_stress_and_displacement_continuity}
 \end{equation}
 where $\textbf{n}_{ab}$ is the unit vector normal to the interface between anode materials $a$ and $b$.
 
 \subsubsection{Chemical Model}\label{subsubsec:Chemical_Model}
 
 We model the movement of lithium within the anode as a diffusive process, with flux $\textbf{j}^*$ given by
 \begin{equation}
 \textbf{j}^* = -\frac{D_a}{R_gT}c_a^*\del^*\mu_a^* \quad \text{ in } \Omega_a,
 \end{equation}
 where $\mu_a^*$ is the chemical potential of the lithium intercalated into material $a$, $D_a$ is the diffusion coefficient of lithium through material $a$, $R_g$ is the ideal gas constant 8.314 J mol$^{-1}$ K$^{-1}$ and $T$ is the temperature, which we assume to be constant, uniform and equal to 298 K.
 The time-scale we focus on is much slower than any dynamics in the system, so that we may make the quasi-steady approximation $j^* = 0$.
 Thus
 \begin{equation}
 c_a^*\del^*\mu^*_a = 0  \quad\quad\quad \text{ in }\Omega_a, \label{eq:dimensional_chemical_equilibrium}
 \end{equation}
 implying the chemical potential of the lithium is uniform in each anode material for all non-trivial concentrations $c_a$.
 At the interface with the electrolyte we have
 \begin{equation}
 [\mu^*_a]^+_- = 0 \quad \text{ on } \Gamma_{e}. \label{eq:dimensional_potential_continuity_e}
 \end{equation}
 Similarly, at the interface between anode materials $\mu^*_a$ must be continuous:
 \begin{equation}
 [\mu^*_a]^+_- = 0 \quad \text{ on } \Gamma_{ab}, \label{eq:dimensional_potential_continuity_ab}
 \end{equation}
 and therefore $\mu^*_a$ takes the same value for $a = 1, \dots n$ and we denote this value by $\mu^*$.
 
 We define the chemical potential of the lithium intercalated into material $a$ using the stress-dependent chemical potential commonly used with linear elasticity \cite{Larche1973a}, given by
 \begin{equation}
 \mu^* = \tilde{\mu}_a^{\text{SF},*}(c_a^*) - \eta_a V_a^m \tr\big(\bs{\sigma}^*\big) \quad \text{ in } \Omega_a, \label{eq:dimensional_potential}
 \end{equation}
 where, $\text{tr}(\bs{\sigma}^*)$ denotes the trace of $\bs{\sigma}^*$ and $\tilde{\mu}_a^\text{SF,*}(c_a^*)$ is the stress-free chemical potential of lithium in material $a$.
 We explain how $\tilde{\mu}_a^\text{SF,*}$ is obtained in Section~\ref{subsec:using_OCVs}.
 
 To close the system, we prescribe the total amount of intercalated lithium.
 We describe this as a proportion of the maximum amount possible using the state of charge parameter $c_0 \in [0,1]$, where $c_0=0$ denotes no lithium and $c_0=1$ denotes a fully lithiated anode.
 Thus we impose
 \begin{equation}
 c_0\sum_{a=1}^n \bigg(\int_{\Omega_a} c_a^\text{max} \;\; \text{d}V\bigg) = \sum_{a=1}^n \bigg(\int_{\Omega_a} c_a^* \;\; \text{d}V\bigg),\label{eq:dimensional_total_amount_of_lithium}
 \end{equation}
 where $c_a^\text{max}$ is the maximum lithium concentration possible in material $a$.
 
 In summary, the full dimensional mechanical model is given by the governing equation \eqref{eq:dimensional_mechanical_equilibrium}, the Cauchy stress \eqref{eq:dimensional_stress} and the boundary conditions \eqref{eq:dimensional_traction_free}-\eqref{eq:dimensional_stress_and_displacement_continuity}.
 The full dimensional chemical model is given by the uniform chemical potential \eqref{eq:dimensional_potential} and state of charge condition \eqref{eq:dimensional_total_amount_of_lithium}.
 Given a value of $c_0 \in [0,1]$, the concentration and displacement profile can be calculated from the model.
 
  \subsection{Using OCVs to Calculate Chemical Potential}\label{subsec:using_OCVs}
 
 In this section, we outline how the stress-free chemical potential of the lithium atoms $\tilde{\mu}_a^{\text{SF},*}(c_a^*)$ in a single anode material $a$ can be determined from the OCV of that material.
 We follow Bazant \cite{bazant2013theory} and Newman \textit{et al.} \cite{newman2012electrochemical} by equating the electrochemical potentials of the reactants and the products of the surface reaction occurring at $\Gamma_e$.
 The reversible reaction at the surface of each electrode during charging or discharging is
 \begin{equation}
 \text{Li}_\text{s} \xleftrightharpoons[]{} \text{Li}^+_\text{aq} + e^-_\text{s}, \label{eq:chemical_reaction}
 \end{equation}
 where the subscript s denotes that the lithium atoms and electrons are in the solid electrode and the subscript aq denotes that the lithium ions are dissolved in the electrolyte.
 
 As these reactions are in equilibrium during OCV measurement, we may equate the total electrochemical potential of the reactants and products.
 Since we are only concerned with single-material electrodes here, and the electrodes are assumed to be unconfined, there is no stress-dependent contribution to the chemical potential.
 The balances of electrochemical potentials at the surface of the anode and the cathode
are thus given by
 \begin{align}
 \tilde{\mu}^{\text{SF},*}_{\text{Li},a}(c_{\text{Li},a}^*) &= \mu_{\text{Li}^+,\text{el}}^*(c_{\text{Li}^+,\text{el}}^*) + ze\phi^*_\text{el} + \mu_{e^-,a}^*(c_{e^-,a}^*) - ze\phi^*_a, \label{eq:chemical_potential_of_reactions1} \\
 \tilde{\mu}^{\text{SF},*}_{\text{Li},c}(c_{\text{Li},c}^*) &= \mu_{\text{Li}^+,\text{el}}^*(c_{\text{Li}^+,\text{el}}^*) + ze\phi^*_\text{el} + \mu_{e^-,c}^*(c_{e^-,c}^*) - ze\phi^*_c,\label{eq:chemical_potential_of_reactions2}
 \end{align}
 respectively, where, $\phi^*$ represents the electrical potential. 
 The first subscript of the chemical potentials $\mu^*$ in \eqref{eq:chemical_potential_of_reactions1}-\eqref{eq:chemical_potential_of_reactions2} denotes the chemical species that the chemical potential corresponds to.
 The subscript of the electrical potentials and the second subscript of the chemical potentials denote the phase which the species is in, $a$ denoting anode, $c$ denoting cathode and el denoting electrolyte.
 We have used the first subscript to avoid confusion with specifying the chemical potentials of different species, however, it should be noted that $\tilde{\mu}^{\text{SF},*}_{a}(c_a^*)$ in the notation of Section~\ref{subsubsec:Chemical_Model} is $\tilde{\mu}^{\text{SF},*}_{\text{Li},a}(c_{\text{Li},a}^*)$ in \eqref{eq:chemical_potential_of_reactions1}.
 Finally, the charge of the Li$^+$ ions is denoted by $z$ and $e$ is the fundamental charge 1.60217$\times 10 ^{-19}$C.

 The OCV, denoted by $E^\text{OC}$, measures the difference in the electrochemical potential of the electrons in the anode and the cathode, divided by the charge of the electron $-e$ \cite{newman2012electrochemical}, and hence is given by
 \begin{equation}
  E^\text{OC} = -\frac{\mu_{e,a}^*(c_{e,a}^*)}{e} + z\phi^*_a + \frac{\mu_{e,c}^*(c_{e,c}^*)}{e} - z\phi^*_c.\label{eq:OCV_definition}
 \end{equation}
 We can thus subtract \eqref{eq:chemical_potential_of_reactions2} from \eqref{eq:chemical_potential_of_reactions1}, substitute \eqref{eq:OCV_definition}, and rearrange to obtain an expression for the chemical potential of the lithium in the anode in terms of the OCV, given by
 \begin{align}
 \tilde{\mu}^{\text{SF},*}_{\text{Li},a}(c_{\text{Li},a}^*) = -eE^\text{OC} + \mu^{\text{SF},*}_{\text{Li},c}(c_{\text{Li},c}^*).\label{eq:potential_from_OCV}
 \end{align}
 Hence, we convert the OCV, $E^\text{OC}$, conventionally measured in volts, to chemical potential $\tilde{\mu}^{\text{SF},*}_{\text{Li},a}$, measured in J mol$^{-1}$, by multiplying by the fundamental charge, $e$ (J V$^{-1}$), and Avagadro's number, 6.02214086 $\times 10^{23}$ mol$^{-1}$.
 
 The chemical potential for the lithium in the cathode $\mu^{\text{SF},*}_{\text{Li},c}(c_{\text{Li},c}^*)$ in \eqref{eq:potential_from_OCV} does not concern this work for two reasons.
 Firstly, the OCV of the anode material is often measured against a cathode of lithium metal, thus the backwards reaction of \eqref{eq:chemical_reaction} is lithium plating and so $\mu^{\text{SF},*}_{\text{Li},c}$ is independent of $c_{\text{Li},c}^*$.
 This makes $\mu^{\text{SF},*}_{\text{Li},c}$ an additive constant, only shifting the reference potential for $\tilde{\mu}^{\text{SF},*}_{\text{Li},a}$.
 Secondly, we are only concerned with comparing the potentials of the lithium in different anode materials or comparing the potentials of lithium in an anode material at different states of charge.
 Therefore, as long as the cathode used when calculating the OCV of the different materials is the same, this additive constant will cancel when comparing them.
 
 \subsection{Nondimensionalisation}\label{subsec:nondimensionalisation}
 
 We now nondimensionalise the variables in our chemo-mechanical model.
 The spatial coordinate $\bs{x}^*$, the displacement vector $\bs{u}^*$, the Cauchy stress $\bs{\sigma}^*$, the concentration in each material $c_a^*$ and the stress-free and stress-dependent chemical potentials $\mu_a^\text{SF,*}$ and $\mu_a^*$ are nondimensionalised by setting
 \begin{equation}
 \begin{aligned}
 \bs{x}^*& = L\bs{x}, \quad\quad\quad \bs{u}^* = L \eta_1 V_1^m c_1^\text{max}\bs{u} \quad\quad\quad \bs{\sigma}^* = G_1^*(0) \eta_1 V^m_1 c_1^\text{max}\bs{\sigma}, \\
 c_a^* = c_a^\text{max} c_a,& \quad\quad\quad \tilde{\mu}_a^{\text{SF},*}(c_a^*) = R_g T \mu_a^\text{SF}(c_a), \quad\quad\quad \mu_a^* = R_g T \mu_a, \quad \text{ for } a = 1\dots n, \label{eq:linear_nondim_factors}
 \end{aligned}
 \end{equation}
 respectively.
 Here, $L$ is the representative length-scale of the anode particle, $G_1(0)$ is the shear modulus of material 1 at zero lithiation and we have chosen to scale each variable using the parameters relevant to material 1.
 We note that the linear forms of the stress \eqref{eq:dimensional_stress} and the chemical potential \eqref{eq:dimensional_potential} are only justified in situations where $\eta_a V_a^m c_a^\text{max} \ll 1$ for $a = 1,\dots,n$, since then the linearisation of the elastic strain tensor in \eqref{eq:linear_elastic_strain_tensor} is permitted.
 A full derivation from the fully nonlinear formulation to the mechanical model we use here can be found in \cite{Roper2019silicon}.
 
 Using this nondimensionalisation, the governing equation \eqref{eq:dimensional_mechanical_equilibrium} is
 \begin{align}
 \del \cdot \bs{\sigma} &= 0 \quad &\text{ in } \Omega, \label{eq:nondimensional_mechanical_equilibrium} 
 \end{align}
 with
 \begin{align}
 \bs{\sigma} &= \mathbb{C}_a : \big(\del\bs{u} - \gamma_ac_a\bs{1}\big) \nonumber \\
 &= \lambda_a(c_a)\kd{ij}\pd{u_k}{x_k} + G_a(c_a)\bigg(\pd{u_i}{x_j} + \pd{u_j}{x_i}\bigg) - \gamma_ac_a(3\lambda_a(c_a)+2G_a(c_a))\kd{ij} \quad \text{ in } \Omega_a, \label{eq:nondimensional_stress}
 \end{align}
 while the uniform chemical potential is given by
 \begin{equation}
 \mu = \mu_a^\text{SF}(c_a) - S_a^d \tr\big(\bs{\sigma}\big) \quad \text{ in } \Omega_a, \label{eq:nondimensional_potential}
 \end{equation}
 where the dimensionless parameters are given by
 \begin{equation}
 \lambda_a(c_a) = \frac{\lambda_a^*(c_a^*)}{G_1^*(0)}, \quad G_a(c_a) = \frac{G_a^*(c_a^*)}{G_1^*(0)}, \quad \gamma_a = \frac{\eta_aV^m_ac_a^\text{max}}{\eta_1 V_1^m c_1^\text{max}}, \quad \text{ and } \quad S_a^d = \frac{\eta_a\eta_1V^m_aV^m_1c_1^\text{max}G_1^*(0)}{R_gT}.\label{eq:nondimensional_parameters}
 \end{equation}
 Henceforth, we drop the explicitly written lithiation-dependence of the Lam\'{e} parameters and write $\lambda_a \equiv \lambda(c_a)$ and $G_a \equiv G_a(c_a)$.
 The boundary conditions are
 \begin{align}
 \bs{\sigma}\cdot\textbf{n} &= 0 \quad\quad\quad \text{ on } \Gamma_e,\label{eq:nondimensional_traction_free} \\ 
 [\bs{\sigma} \cdot \textbf{n}]^+_- = [\bs{u}]^+_- &= 0 \quad\quad\quad \text{ on }\Gamma_{ab}, \label{eq:nondimensional_stress_and_displacement_continuity}
 \end{align}
 while the SOC condition is
 \begin{equation}
 c_0\sum_{a=1}^n \bigg(\int_{\Omega_a} c_a^\text{max} \;\; \text{d}V\bigg) = \sum_{a=1}^n \bigg(\frac{c_a^\text{max}}{c_1^\text{max}}\int_{\Omega_a} c_a \;\; \text{d}V\bigg). \label{eq:nondimensional_total_amount_of_lithium}
 \end{equation}
 
 \subsection{Spherical Geometry}\label{subsec:spherical_geometry}
 
 We now focus on the specific geometry of a single, spherical, radially symmetric nano-particle comprising a core of anode material 1 enclosed in a shell of anode material 2.
 We denote the radius of the core made of material 1 by $R_1$, and thus material 1 occupies the domain $\Omega_1 = \{r | 0\leq r<R_1\}$.
 We denote the radius of the entire nano-particle by $R_2$ so that the shell made of material 2 has thickness $R_2-R_1$ and occupies the domain $\Omega_2 = \{r | R_1<r\leq R_2\}$.
 This geometry is shown schematically in Figure~\ref{fig:dimensionless_OCV_curves}.
 Our representative length-scale of the anode particle $L$ is now given by the radius, $R_2$.
 
 \begin{figure}[b!]
 \centering
 \includegraphics[scale=2.0]{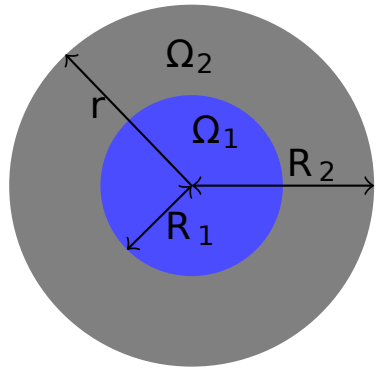}
 \caption{Schematic of a slice through a spherical nano-particle with a spherical central core of material 1 denoted by $\Omega_1$ and a shell of material 2 denoted by $\Omega_2$. The outer radii of the two regions are labelled as $R_1$ and $R_2$, respectively, and the direction of the spatial variable $r$ is labelled.}
 \label{fig:spherical_geometry}
 \end{figure}
  
 We assume the displacement $\bs{u}$ is solely radial due to symmetry and thus we write $\bs{u} = u(r)\bs{e}_r$.
 The Cauchy stress and the chemical potential of lithium in material $a$, \eqref{eq:nondimensional_stress} and \eqref{eq:nondimensional_potential}, are then given by
 \begin{align}
 \bs{\sigma} &= 
 \text{ diag}
 \begin{bmatrix}
 \lambda_a\Big(\fd{u}{r} + \frac{2u}{r}\Big) + 2G_a\fd{u}{r} - \gamma_a(3\lambda_a+2G_a) c_a \\
 \lambda_a\Big(\fd{u}{r} + \frac{2u}{r}\Big) + 2G_a\frac{u}{r} - \gamma_a(3\lambda_a+2G_a)c_a \\
 \lambda_a\Big(\fd{u}{r} + \frac{2u}{r}\Big) + 2G_a\frac{u}{r} - \gamma_a(3\lambda_a+2G_a)c_a
 \end{bmatrix} \quad &\text{ in } \Omega_a, \label{eq:linear_stress_tensor} \\
  \mu &= \mu_a^\text{SF}\big(c_a(r)\big) - S_a^d(3\lambda_a+2G_a)\bigg(\frac{1}{r^2}\fd{}{r}\big(r^2u\big) - 3\gamma_ac_a\bigg) \quad &\text{ in } \Omega_a,\label{eq:chemical_potential_radial}
 \end{align}
 respectively. 
 The mechanical equilibrium equation \eqref{eq:nondimensional_mechanical_equilibrium} becomes
 \begin{align}
 &\fd{}{r}\bigg(\frac{1}{r^2}\fd{}{r}(r^2u)\bigg) - \frac{\gamma_a(3\lambda_a+2G_a)}{\lambda_a+2G_a}\fd{c_a}{r} = 0,\quad &\text{ in } \Omega_a, \label{eq:1D_linear_momentum_balance_disp}
 \end{align}
  
 Integrating \eqref{eq:1D_linear_momentum_balance_disp} gives
 \begin{equation}
 \fd{u}{r} + \frac{2u}{r} - \frac{\gamma_a(3\lambda_a+2G_a)}{\lambda_a+2G_a}c_a = \beta_a \quad \text{ in } \Omega_a, \label{eq:uniform_c_derivation_one_integration}
 \end{equation}
 where the integration constant $\beta_a$ may be different in the two materials.
 The trace of the Cauchy stress tensor \eqref{eq:linear_stress_tensor} is given by
 \begin{equation}
 \tr(\bs{\sigma}) = \big(3\lambda_a+2G_a\big)\bigg(\fd{u}{r} + \frac{2u}{r} - 3\gamma_ac_a\bigg) \quad \text{ in } \Omega_a. \label{eq:trace_of_Cauchy_stress}
 \end{equation}
 Using \eqref{eq:uniform_c_derivation_one_integration} gives
 \begin{equation}
 \tr(\bs{\sigma}) = (3\lambda_a+2G_a)\bigg(\beta_a +  \frac{\gamma_a(3\lambda_a+2G_a)}{\lambda_a+2G_a}c_a - 3\gamma_ac_a\bigg) \quad \text{ in } \Omega_a.\label{eq:trace}
 \end{equation}
 Thus the trace of the Cauchy stress and therefore the chemical potential, \eqref{eq:chemical_potential_radial}, can be written in terms of $c_a$ only.
 Since $\mu_a$ is uniform in each material we conclude that $c_a$ is also uniform in each material.
 
 Substituting a uniform $c_a$ into \eqref{eq:uniform_c_derivation_one_integration} and solving the resulting ODE yields
 \begin{equation}
 u = A_a r + \frac{B_a}{r^2} \quad \text{ in } \Omega_a, \label{eq:linear_elasticity_solution}
 \end{equation}
 where $A_a = 3\beta_a + (3\gamma_a(3\lambda_a+2G_a)c_a)/(\lambda_a+2G_a)$ is now the constant arising from the first integration, $B_a$ is the constant arising from the second integration and each integration constant may be different in the two materials.
 Substituting \eqref{eq:linear_elasticity_solution} into \eqref{eq:chemical_potential_radial} gives
 \begin{equation}
 \mu = \mu^\text{SF}_a(c_a) - 3S^d_a\big(3\lambda_a + 2G_a\big)(A_a - \gamma_ac_a), \quad \text{ in } \Omega_a. \label{eq:linear_potential_with_trace_expression_final}
 \end{equation}
 We can substitute \eqref{eq:linear_elasticity_solution} into \eqref{eq:linear_stress_tensor} and write the radial stress and hoop stresses as
 \begin{align}
 \sigma_{rr} &= (3\lambda_a + 2G_a)(A_a-\gamma_ac_a) - \frac{4G_aB_a}{r^3} \quad \text{ in } \Omega_a, \label{eq:linear_1D_general_radial_stress}\\
 \sigma_{\theta\theta} = \sigma_{\phi\phi} &= (3\lambda_a + 2G_a)(A_a-\gamma_ac_a) + \frac{2G_aB_a}{r^3} \quad \text{ in } \Omega_a, \label{eq:linear_1D_general_hoop_stress}
 \end{align}
 respectively.
 It remains to use the boundary conditions to determine the integration constants $A_a$ and $B_a$.
 
 The boundary conditions \eqref{eq:nondimensional_traction_free} and \eqref{eq:nondimensional_stress_and_displacement_continuity} for the spherical geometry are written as
  \begin{align}
 \sigma_{rr} &= 0, \quad \text{ on } r=1, \label{eq:zero_traction_radial}\\
 [\sigma_{rr}]^+_- = [u]^+_- &= 0, \quad \text{ on } r=\rad, \label{eq:disp_and_stress_continuity_radial}
 \end{align}
 respectively, where $R = R_1/R_2$.
 Additionally, we ensure that the displacement is bounded at the origin by prescribing
 \begin{equation}
 u=0 \text{ at } r=0. \label{eq:finite_displacement_at_r=0}
 \end{equation}
 Substituting \eqref{eq:linear_elasticity_solution} and \eqref{eq:linear_1D_general_radial_stress} into \eqref{eq:zero_traction_radial}-\eqref{eq:finite_displacement_at_r=0} gives
 \begin{align}
 A_1 &= \frac{1}{\omega}\Big[(3\lambda_1+2G_1)\big((3\lambda_2+2G_2) + 4G_2 \rad^3\big)\gamma_1 c_1 + 4G_2(1-\rad^3)(3\lambda_2+2G_2)\gamma_2 c_2\Big],
\label{eq:linear_spherical_A1}\\
 B_1 &= 0, \\
 A_2 &= \frac{1}{\omega}\Big[(3\lambda_2+2G_2)\big(4G_2 (1-\rad^3) + (3\lambda_1+2G_1)\big)\gamma_2 c_2 + 4G_2 \rad^3(3\lambda_1 + 2G_1)\gamma_1 c_1 \Big], \label{eq:linear_spherical_A2}\\
 B_2 &= \frac{1}{\omega}\Big[(3\lambda_1+2G_1)(3\lambda_2+2G_2)(\gamma_1 c_1 - \gamma_2 c_2)\rad^3\Big],\label{eq:linear_spherical_B2}
 \end{align}
 where
 \begin{equation}
 \omega = (3\lambda_1+2G_1)(3\lambda_2+2G_2) + 4G_2\big((3\lambda_2+2G_2)(1-\rad^3) + (3\lambda_1+2G_1)\rad^3\big).
 \label{eq:linear_spherical_omega}
 \end{equation}
 Thus the displacements and stress are given once the concentrations $c_1$ and $c_2$ have been determined.
 
 The SOC condition \eqref{eq:nondimensional_total_amount_of_lithium} becomes
 \begin{equation}
 c_0\bigg[\rad^3 + \big(1-\rad^3\big)\frac{c_2^\text{max}}{c_1^\text{max}}\bigg] = \rad^3 c_1 + \frac{c_2^\text{max}}{c_1^\text{max}}(1-\rad^3) c_2, \label{eq:total_Li_condition_spherical}
 \end{equation}
 which we can rearrange as
 \begin{equation}
 c_1 = c_0 + \frac{c_2^\text{max}}{c_1^\text{max}}\big(1-\rad^{-3}\big)\big(c_2-c_0\big). \label{eq:c1_in_terms_of_c2}
 \end{equation}
 Finally, we use \eqref{eq:linear_potential_with_trace_expression_final} to relate the concentrations $c_1$ and $c_2$.
 Combining this with \eqref{eq:c1_in_terms_of_c2} gives a single algebraic equation in $c_2$ for a given $c_0$ to be solved numerically\footnote{For very high values of $\rad$ and very low values of $c_0$, this function has multiple roots. For consistency, we take the lowest root in these cases.}.
 As the concentrations $c_a$ must be between zero and one due to the nondimensionalisation \eqref{eq:linear_nondim_factors}, we place bounds on $c_2$, ensuring that both $c_1, c_2 \in [0,1]$.
 We use \eqref{eq:c1_in_terms_of_c2} to find that
 \begin{equation}
 \max\bigg\{0, \frac{(1-c_0)c^\text{max}_1 \rad^3}{c^\text{max}_2(\rad^3-1)} + c_0\bigg\} < c_2 < \min\bigg\{1, c_0 - \frac{c_0c_1^\text{max}\rad^3}{c_2^\text{max}(\rad^3-1)}\bigg\},
 \end{equation}
 where the first and second elements of the maximum and minimum functions ensure $c_2 \in [0,1]$ and $c_1\in[0,1]$, respectively.

\section{Chemical Results}\label{sec:Chemical_results}
 
 We illustrate the behaviour of a nano-particle consisting of a silicon core and a graphite shell described by the model and geometry explained in Section~\ref{subsec:spherical_geometry}.
 The relevant mechanical and chemical parameters for these materials are given in Table~\ref{table:CS_parameters}, including the nondimensional parameters $\gamma_a$ and $S^d_a$.
 We calculate the Lam\'{e} parameters of silicon and graphite from the Young's modulus and the Poisson's ratio.
 The MATLAB code to produce the figures in this section is available publicly on GitHub \cite{roper2019code}.
 
 We assume the Young's modulus varies linearly with the lithiation state using the values at zero and full lithiation as reference values.
 Therefore, the dimensional Lam\'{e} parameters are given by
 \begin{equation}
 \lambda_a^* = \frac{E^{0*}_a(1+\eta^E_a V_a^m c_a^*)\nu_a}{(1+\nu_a)(1-2\nu_a)}, \quad \quad G_a^* = \frac{E^{0*}_a(1+\eta^E_a V_a^m c_a^*)}{2(1+\nu_a)}, 
 \end{equation}
 where $E^{0*}_a$ is the Young's modulus of material $a$ at zero lithiation, $\nu_a$ is the Poisson's ratio of material $a$ and $\eta^E_a$ is the linear variation of the Young's modulus of material $a$.
 As the Poisson's ratios of these materials vary negligibly with lithiation state \cite{cui2012finite}, we assume $\nu_a$ to be independent of $c_a$.
 
 \begin{figure}[t!]
 \begin{center}
 \begin{tabular}{ |c|c|c|c| } 
 \hline
 Parameter & Description & Silicon & Graphite \\
 \hline & & & \\
  & & & \vspace{-8mm} \\
 $J^c$ & Relative expanded volume at $c_a=1$ & $3.8^\text{\cite{liu2012size}}$ & $1.1^\text{\cite{qi2014lithium}}$ \\
 $x^\text{max}$ & Maximum stoichiometric ratio of Li & 3.75$^\text{\cite{Hatchard2004}}$ & 0.167$^\text{\cite{chakraborty2015combining}}$ \\ 
 $V^m$ & Molar volume (m$^3$mol$^{-1}$) & $1.205\times 10^{-5\text{\cite{cui2012finite}}}$ & $8.69 \times 10^{-6\text{\cite{yu1999determination}}}$ \\
 $c^\text{max}$ & Maximum Li concentration (mol m$^{-3}$) & $3.11 \times 10^{5\text{\cite{Roper2019silicon}}}$ & $1.92 \times 10^{4\text{\cite{Roper2019silicon}}}$ \\
 $\eta$ & Coefficient of compositional expansion & 0.2489$^\text{\cite{Roper2019silicon}}$ & 0.2$^\text{\cite{Roper2019silicon}}$ \\
 $\nu$ & Poisson's ratio & 0.29$^\text{\cite{qi2014lithium}}$ & 0.32$^\text{\cite{qi2014lithium}}$ \\
 $E_0^*$ & Young's modulus at $c_a = 0$ (GPa)& 96.0$^\text{\cite{qi2014lithium}}$ & 32.0$^\text{\cite{qi2014lithium}}$ \\
 $\eta_a^E$ & Linear conc. dependence of Young's Modulus & $-0.1302^1$ & $14.4375^1$ \\
 $\gamma$ & Nondim. expansion coefficient (Eq. \ref{eq:nondimensional_parameters}) & 1.0 & 0.0357 \\
 $S^d$ & Stress-assisted diffusion param. (Eq. \ref{eq:nondimensional_parameters}) &  42.046 & 1.502 \\
 \hline 
 \end{tabular}\\
 $^1$ Calculated from the Young's modulus at $c_a = 1$ from \cite{qi2014lithium}
 \captionof{table}{Material parameters of silicon and graphite.}
 \label{table:CS_parameters}
 \end{center}
 \end{figure}
 
\begin{figure}[t!]
 \centering
 \includegraphics[scale=0.79]{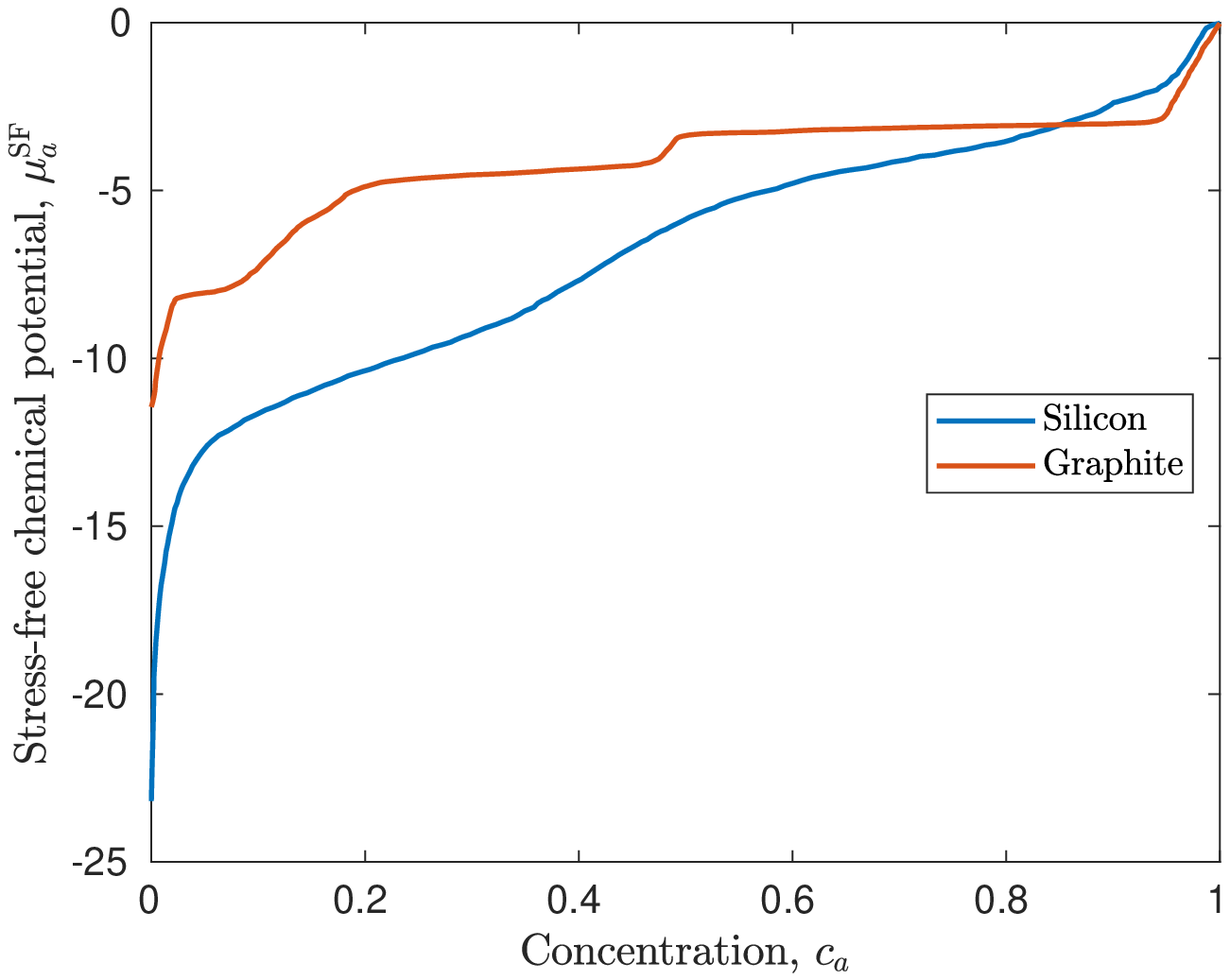}
\caption{Nondimensional stress-free chemical potential $\mu^{\text{SF}}_a$ of the lithium in each material as a function of nondimensional concentration of lithium for silicon and graphite, $c_1$ and $c_2$.
 Data captured with permission using grabit software for MATLAB from Figure 4c in \cite{li2007situ} (silicon) and Figure 1 in \cite{Reynier2003a} (graphite), and linearly interpolated.}
 \label{fig:dimensionless_OCV_curves}
 \end{figure}

 We plot the non-dimensionalised stress-free chemical potentials of the lithium intercalated into silicon and graphite, $\mu_1^\text{SF}$ and $\mu_2^\text{SF}$, in Figure~\ref{fig:dimensionless_OCV_curves}.
 It can be seen that there are several regions where a small change in lithium concentration causes a large change in the chemical potential, as noted in \cite{Birkl2015a}.
 
\begin{figure}[t!]
 \centering
 \includegraphics[scale=0.79]{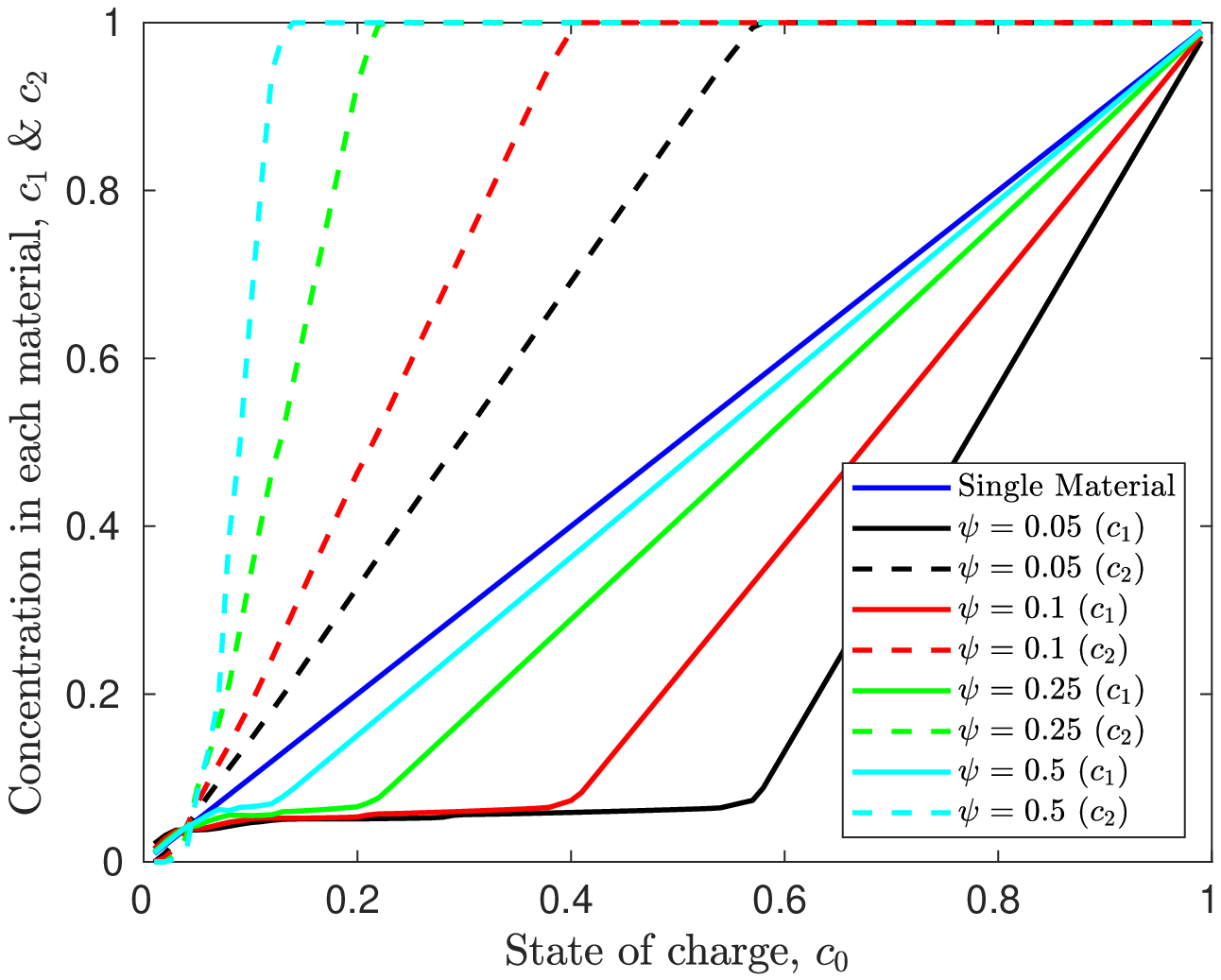}
 \caption{Nondimensional lithium concentrations within each material, $c_1$ (solid lines) and $c_2$ (dashed lines), against state of charge, $c_0$, for different core volume fractions, $\volfrac$. We also plot $c_1 = c_2 = c_0$ labelled as ``Single Material'' to highlight the difference in concentration caused by the presence of a second material.}
 \label{fig:spherical_linear_conc_vs_c0}
 \end{figure}

 In Figure~\ref{fig:spherical_linear_conc_vs_c0} we plot the lithium concentration in each material, $c_1$ and $c_2$, against $c_0$ for different values of the volume fraction $\volfrac = \rad^3$ of the silicon core.
 It can be seen that the graphite is saturated ($c_2 = 1$) at fairly low $c_0$ values, especially for large silicon cores, whereas the lithium concentration in the silicon, $c_1$, remains relatively low until the graphite is saturated.
 After this point, all the lithium being intercalated must be intercalated into the silicon and thus $c_1$ is linear in $c_0$.
 
 \begin{figure}[t!]
 \centering
 \includegraphics[scale=0.79]{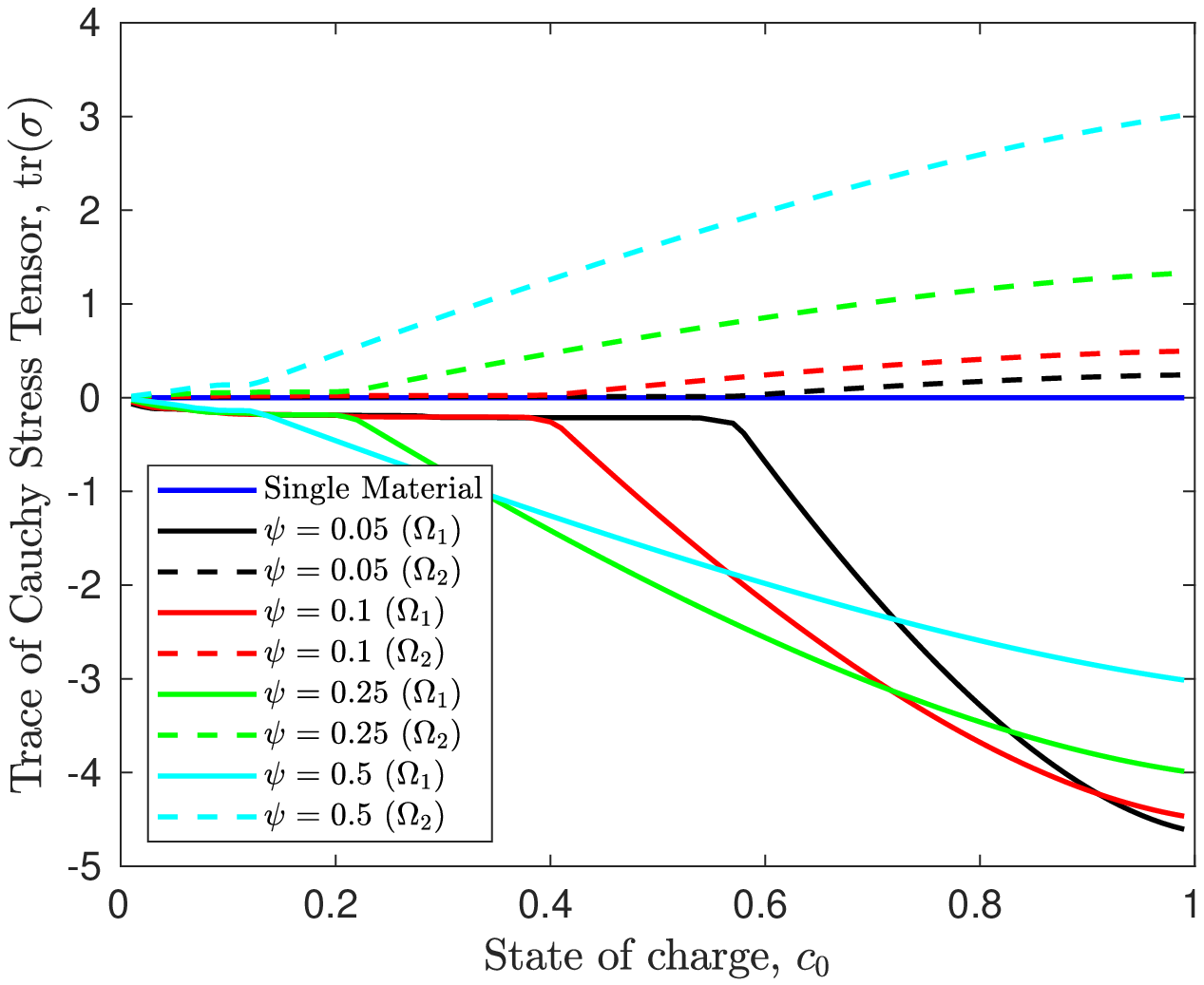}
 \caption{Nondimensional trace of the Cauchy stress tensor, $\text{tr}(\bs{\sigma})$ for each material (solid lines for material 1, dashed lines for material 2), against state of charge, $c_0$, and for different core volume fractions, $\volfrac$.}
 \label{fig:trace}
 \end{figure}
 
 The saturation of the graphite, and the low $c_1$ values, at low SOC are due to the hydrostatic stresses induced in the silicon core and graphite shell.
 The trace of the Cauchy stress in each material, given by \eqref{eq:trace}, is plotted against $c_0$ in Figure~\ref{fig:trace} for different silicon volume fractions, $\volfrac$.
 Since $\gamma_1 \gg \gamma_2$, the expansion of the silicon is much greater than that of the graphite.
 Therefore, when lithiated, the expansion of the silicon is being constrained by the graphite shell, inducing a compressive stress in the silicon, causing $\text{tr}(\bs{\sigma})$ to be negative in $\Omega_1$.
 Conversely, the graphite is being stretched by the large expansion of silicon, inducing a tensile stress, causing $\text{tr}(\bs{\sigma})$ to be positive in $\Omega_2$.
 The tensile stress in the graphite lowers the chemical potential of the lithium whereas the compressive stress in the silicon increases the chemical potential of the lithium.
 Therefore, for the chemical potentials of the lithium in each material to be equal, the stress-free potential of the lithium in the silicon must therefore decrease and the stress-free potential of the lithium in the graphite must increase.
 The stress-free potential of the lithium in each material is a monotonically increasing function of the concentration, therefore $c_1$ must decrease and $c_2$ must increase.
 
 \begin{figure}[t!]
 \centering
\includegraphics[scale=0.79]{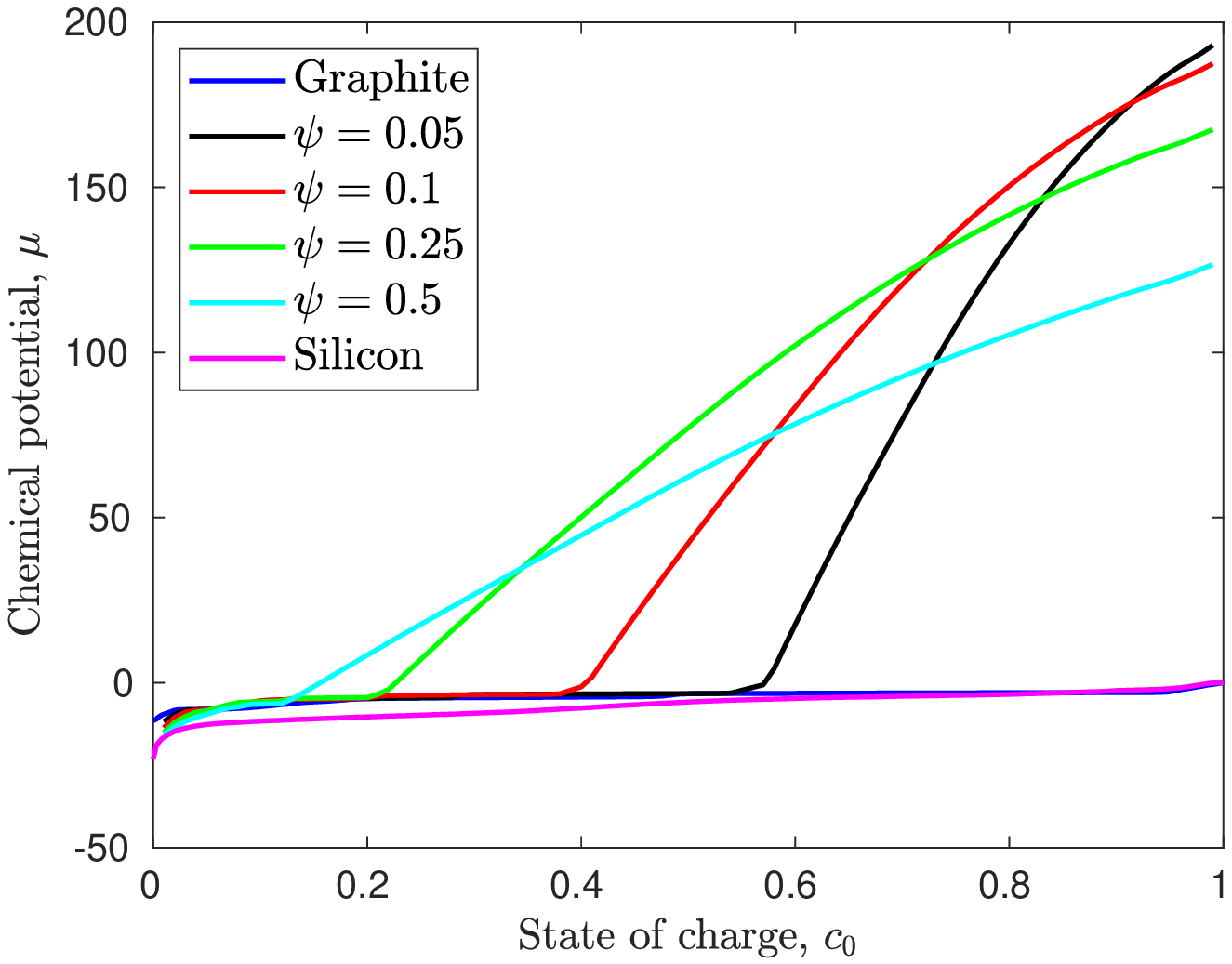}
 \caption{Nondimensional chemical potential, $\mu$, against state of charge, $c_0$, for different core volume fractions, $\volfrac$. The chemical potentials of lithium in graphite and silicon from Figure~\ref{fig:dimensionless_OCV_curves} are also plotted as `Graphite' and `Silicon', respectively.}
 \label{fig:spherical_linear_potential_vs_c0}
 \end{figure}
 
 The low $c_1$ values are present for a larger range of SOC for small silicon cores than for large ones since a thicker graphite shell is stronger, inducing a larger compressive stress on the silicon.
 Likewise, a small silicon core does not stretch the graphite shell as much and thus, for example, the $c_2$ plot for $\volfrac = 0.05$ is more similar to the single material case than that for $\volfrac = 0.5$.
 
 In Figure~\ref{fig:spherical_linear_potential_vs_c0} we plot the chemical potential of the lithium in the hybrid nano-particle against SOC for different values of $\volfrac$.
 It can be seen that for SOC values for which the graphite is saturated, the chemical potential is much greater than that of the individual materials.
 This is because if the graphite is saturated but more lithium is intercalated, the lithium concentration in the silicon must increase.
 This increases $\mu_a^\text{SF}$ but also decreases $\text{tr}(\bs{\sigma})$ further as seen in Figure~\ref{fig:trace}, causing the chemical potential to increase rapidly with increasing SOC.
 
\begin{figure}[t!]
 \centering
 \includegraphics[scale=0.79]{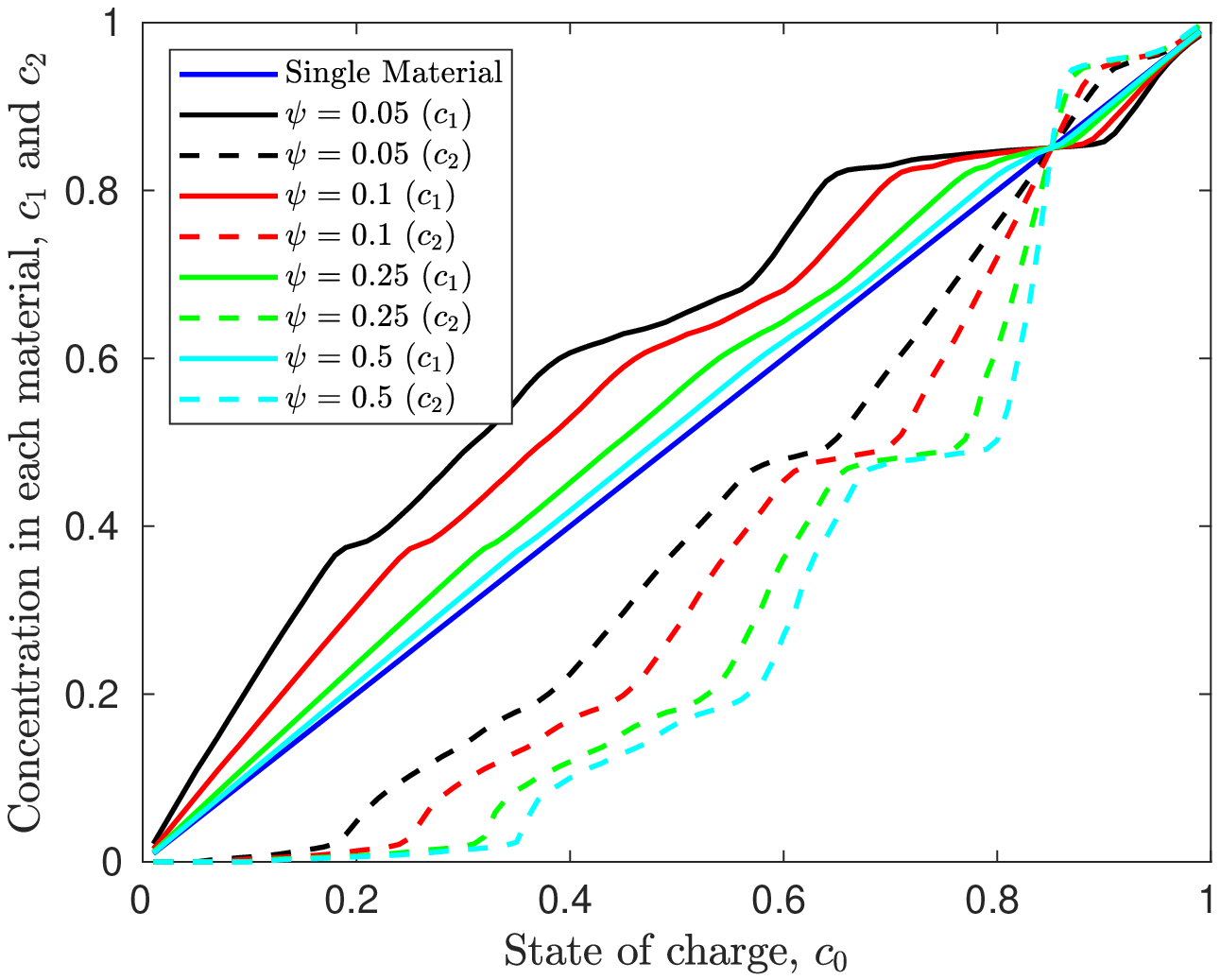}
 \caption{Nondimensional lithium concentrations within each material, $c_1$ (solid lines) and $c_2$ (dashed lines), against state of charge, $c_0$, for different core volume fractions, $\volfrac$, without stress-assisted diffusion ($S^d_a = 0$).}
 \label{fig:spherical_linear_conc_vs_c0_no_SAD}
 \end{figure}

\begin{figure}[t!]
 \centering
\includegraphics[scale=0.79]{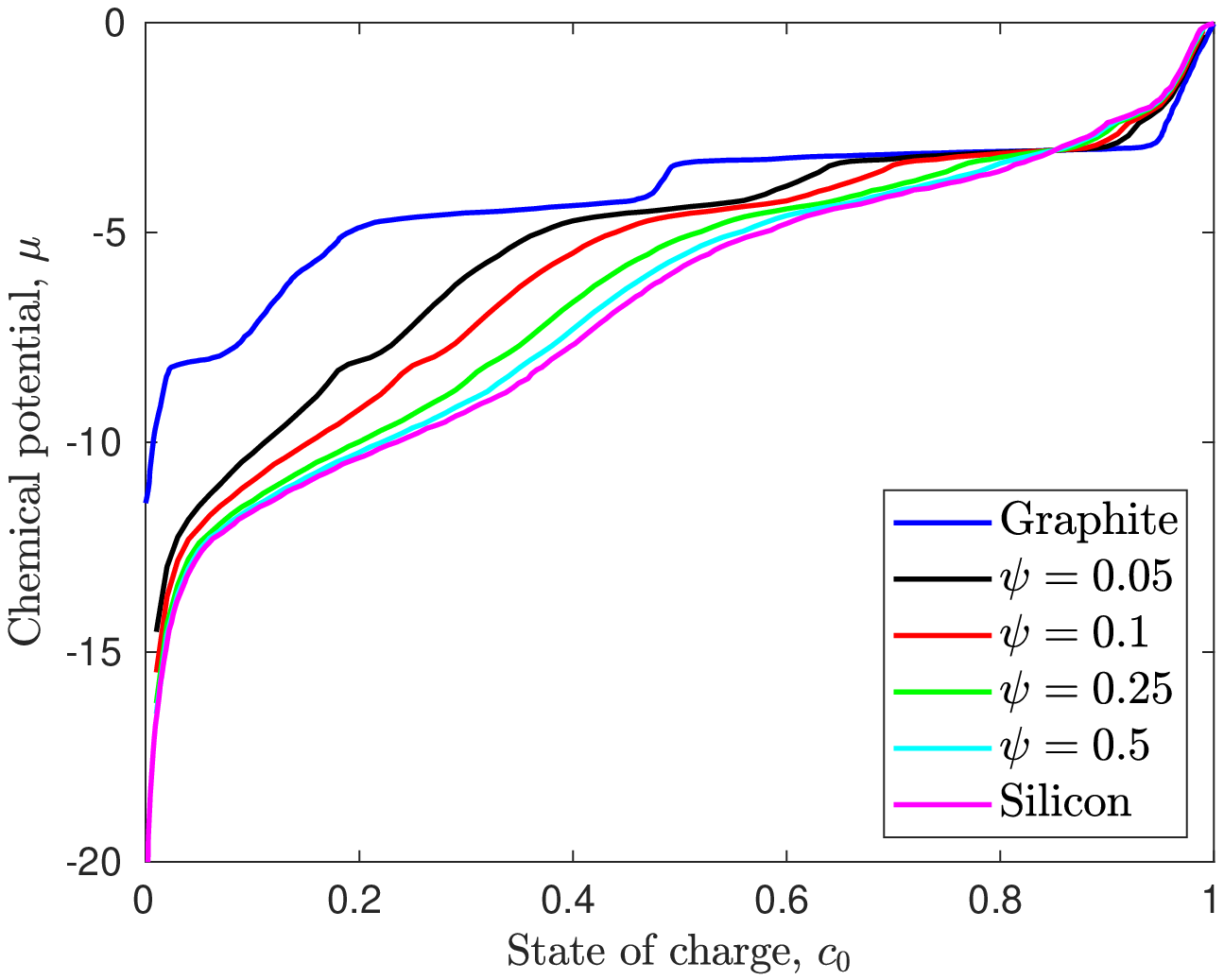}
 \caption{Nondimensional chemical potential, $\mu$, against state of charge, $c_0$, for different core volume fractions, $\volfrac$, without stress-assisted diffusion ($S^d_a = 0$).}
 \label{fig:spherical_linear_potential_vs_c0_no_SAD}
 \end{figure}

 In Figures~\ref{fig:spherical_linear_conc_vs_c0_no_SAD} and \ref{fig:spherical_linear_potential_vs_c0_no_SAD} we replicate Figures~\ref{fig:spherical_linear_conc_vs_c0} and \ref{fig:spherical_linear_potential_vs_c0}, respectively, but we do not account for the effect of stress-assisted diffusion (we set $S_a^d = 0$) to show the importance of including the stress of the anode materials in the chemical potential of the lithium.
 It can be seen in Figure~\ref{fig:spherical_linear_conc_vs_c0_no_SAD} that $c_1>c_2$ for $c_0 \lessapprox 0.8$ when $S^d_a=0$, which is very different to the result found in Figure~\ref{fig:spherical_linear_conc_vs_c0} in which $c_2>c_1$ for $c_0\gtrapprox0.05$.
 This is solely due to the chemical potential of the lithium in the silicon being less than that for graphite for $c_a \lessapprox 0.8$, seen in Figure~\ref{fig:dimensionless_OCV_curves}.
 In Figure~\ref{fig:spherical_linear_potential_vs_c0_no_SAD}, the chemical potential of the lithium in the hybrid nano-particle is approximately an interpolation between the chemical potentials of the lithium within the individual materials, which is also very different to the case in which $S_a^d \neq 0$.
 
 The large differences in $c_1$ and $c_2$ between Figures~\ref{fig:spherical_linear_conc_vs_c0} and \ref{fig:spherical_linear_conc_vs_c0_no_SAD} caused by the exclusion of stress-coupling demonstrates how expansion due to lithium will be dramatically affected by this mechanism.
 Furthermore, the large differences in the chemical potential of the lithium in the hybrid particle between Figures~\ref{fig:spherical_linear_potential_vs_c0} and \ref{fig:spherical_linear_potential_vs_c0_no_SAD} show that stress-modelling is essential in predicting OCVs of hybrid anodes as a function of SOC.
 
 \newpage
 
 \section{Optimal Size of the Silicon Core}\label{sec:optimal_size}
 
 \subsection{Performance Measures}\label{subsec:Performance_measures}
 
 We now discuss the performance measures that might be considered to find an optimal anode design, applying them to the nano-particle geometry shown in Figure~\ref{fig:spherical_geometry}.
 The three properties we consider as these performance measures are i) the total amount of lithium absorbed, ii) the expanded volume of the anode and iii) the maximum stress induced.
 We first derive each of these measures using the results of our model and then investigate how these might be applied in practice to determine the optimal volume fraction of the silicon core in the spherical core--shell nano-particle geometry.
 
 \subsubsection{Amount of Lithium}\label{subsubsec:Amount_of_lithium}
 
 The capacity of a lithium-ion battery is closely related to the amount of lithium that the anode can accommodate and thus this is a very important performance measure of a lithium-ion battery anode.
 For a general anode with domain $\Omega$ and different anode materials $i = 1,\dots, n$, each with domain $\Omega_i$, we measure the total amount of intercalated lithium relative to a fully lithiated anode of material 1 and domain $\Omega$.
 We denote this relative amount of lithium as
 \begin{equation}
 Q = \frac{\sum_{i=1}^n\int_{\Omega_i}c_i^*\;\text{d}V}{\int_{\Omega}c_1^\text{max}\;\text{d}V}.\label{eq:general_Q}
 \end{equation}
 For the core--shell geometry in Figure~\ref{fig:spherical_geometry}, $Q$ is given by
 \begin{equation}
 Q = c_1 \rad^3 + c_2\frac{c_2^\text{max}}{c_1^\text{max}}(1-\rad^3), \label{eq:Q}
 \end{equation}
 where we recall $R=R_1/R_2$.
 
 \subsubsection{Relative Expanded Volume}\label{subsubsec:Relative_Expanded_Volume}
 
 \begin{figure}[b!]
 \centering
 \includegraphics[scale=0.79]{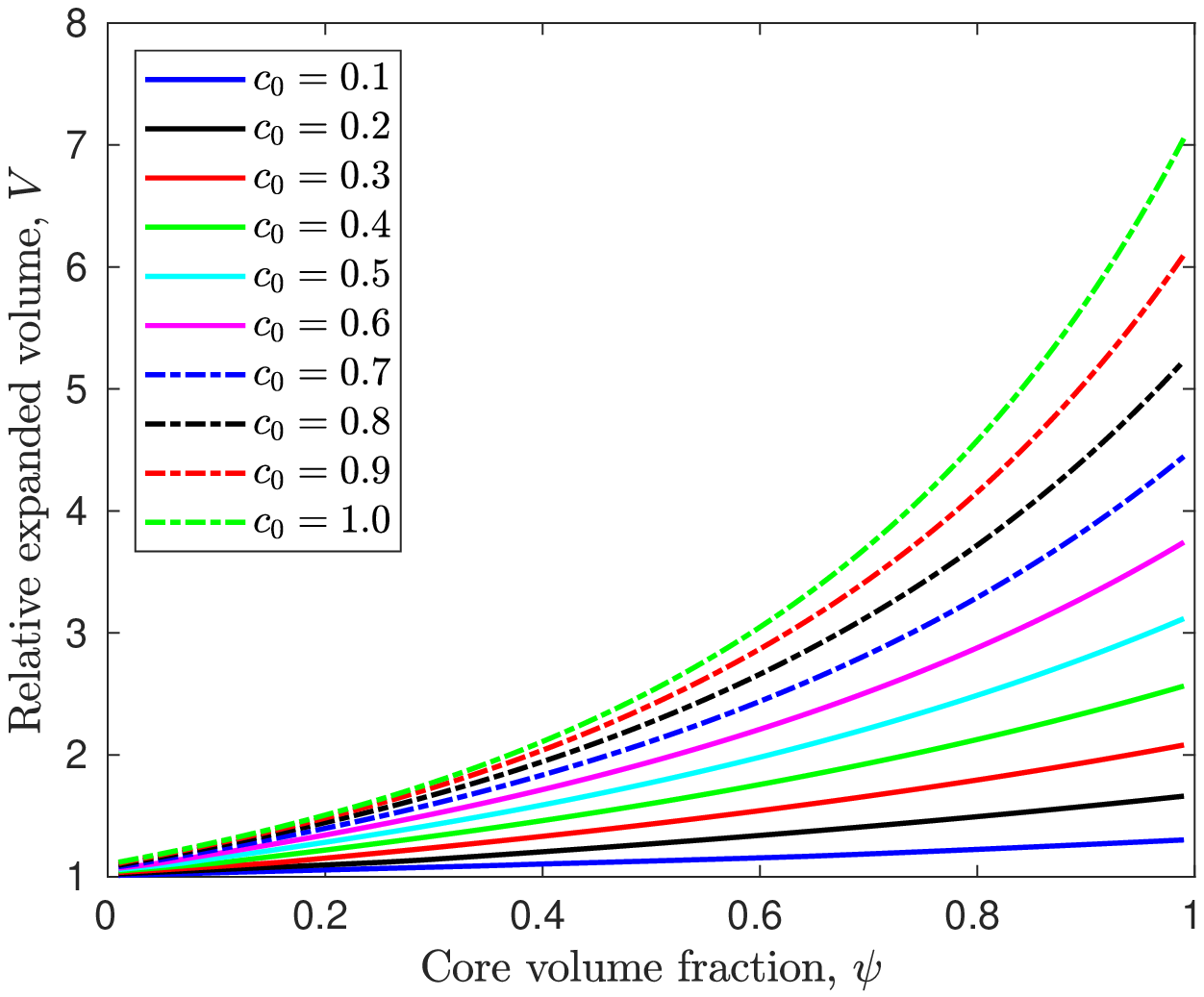}
 \caption{Relative expanded volume, $\expvol$, given by \eqref{eq:spherical_V/V_0}, against core volume fraction, $\volfrac$, for different states of charge, $c_0$.}
 \label{fig:spherical_linear_V/V_0}
 \end{figure}
 
 The expansion of the anode can have serious adverse effects on the battery performance and so we consider this expansion by calculating the volume of the expanded nano-particle compared to the original volume, denoting this as the relative expanded volume, $\expvol$.
 For a general geometry of anode, $\expvol$ would usually have to be calculated numerically, however, for the spherical nano-particle geometry in Figure~\ref{fig:spherical_geometry}, $\expvol$ is given by
 \begin{equation}
 V = \frac{(R_2+u^*(R_2))^3}{R_2^3} = (1+\eta_1V_1^mc_1^\text{max}u(1))^3. \label{eq:spherical_V/V_0}
 \end{equation}
 Here, we use the dimensional displacement $u^*(R_2) = \eta_1V_1^mc_1^\text{max} R_2 u(1)$ as in \eqref{eq:linear_nondim_factors}.
 
 In Figure~\ref{fig:spherical_linear_V/V_0}, we plot $\expvol$ against $\volfrac$ for different states of charge.
 It can be seen that $\expvol$ increases with the volume of the silicon core with this effect becoming more prominent for larger $\volfrac$.
 Additionally, a greater SOC causes a greater $\expvol$ for all volumes of silicon core.
 The calculated $\expvol$ value for a fully lithiated nano-particle with a large silicon core, is much greater than the observed expanded volume for a solely silicon nano-particle, given by $\expvol = J^c_1 = 3.8$.
 This over expansion for large silicon cores is due to our adoption of linear elasticity in Section~\ref{subsubsec:Mechanical_model}.
 This assumes that $\eta_1 V_1^m c_1^\text{max}\ll 1$, whereas for silicon, $\eta_1 V_1^m c_1^\text{max} = 0.933375 \sim 1$.
 Thus the nonlinear elasticity formulation should be retained to produce quantitatively accurate results for silicon at high concentrations.
 
 \subsubsection{Maximum Induced Stress}\label{subsubec:Effective_Stress}
 
 Capacity fade exhibited by expanding anode materials after cycling is often attributed to cracks in the anode material caused by high stresses.
 Thus the final performance measure we use to optimise anode performance is the maximum induced stress.
 We use the von Mises stress \cite{howell2009applied} as a scalar effective stress measure.
 For a general anode geometry, this is given by
 \begin{equation}
 \sigma_\text{eff}^* = \bigg(\frac{(\sigma_{11}^* - \sigma_{22}^*)^2 + (\sigma_{22}^* - \sigma_{33}^*)^2 + (\sigma_{33}^*-\sigma_{11}^*)^2 + 6(\sigma_{12}^{*2} + \sigma_{23}^{*2} + \sigma_{31}^{*2})}{2}\bigg)^\frac{1}{2}.\label{eq:general_von_mises}
 \end{equation}
 In radial symmetry, the von Mises stress \eqref{eq:general_von_mises} can be written as
 \begin{equation}
 \sigma^*_\text{eff} = |\sigma_{rr}^* - \sigma_{\theta\theta}^*|,
 \end{equation}
 so that from \eqref{eq:linear_nondim_factors}, \eqref{eq:linear_1D_general_radial_stress} and \eqref{eq:linear_1D_general_hoop_stress}, we have
 \begin{equation}
 \sigma^*_\text{eff} = \frac{6G_1^* \eta_1 V^m_1 c_1^\text{max}G_a|B_a|}{r^3},
 \end{equation}
 for $a = 1,2$.
 The condition at $r=0$ \eqref{eq:finite_displacement_at_r=0} gives $B_1 = 0$.
 Therefore, the effective stress is only non-zero in $\Omega_2$.
 The $r^{-3}$ dependence shows that this effective stress is the greatest at the minimum value of $r$ in $\Omega_2$, which is $r = \rad = R_1/R_2$.
 Therefore, the maximum induced stress is given by
 \begin{equation}
 \sigma^*_\text{eff}(R_1) = \frac{6\eta_1 V^m_1 c_1^\text{max} G_2^*|B_2|}{\rad^3}.\label{eq:effective_interfacial_stress}
 \end{equation}
 
\begin{figure}[b!]
 \centering
 \includegraphics[scale=0.79]{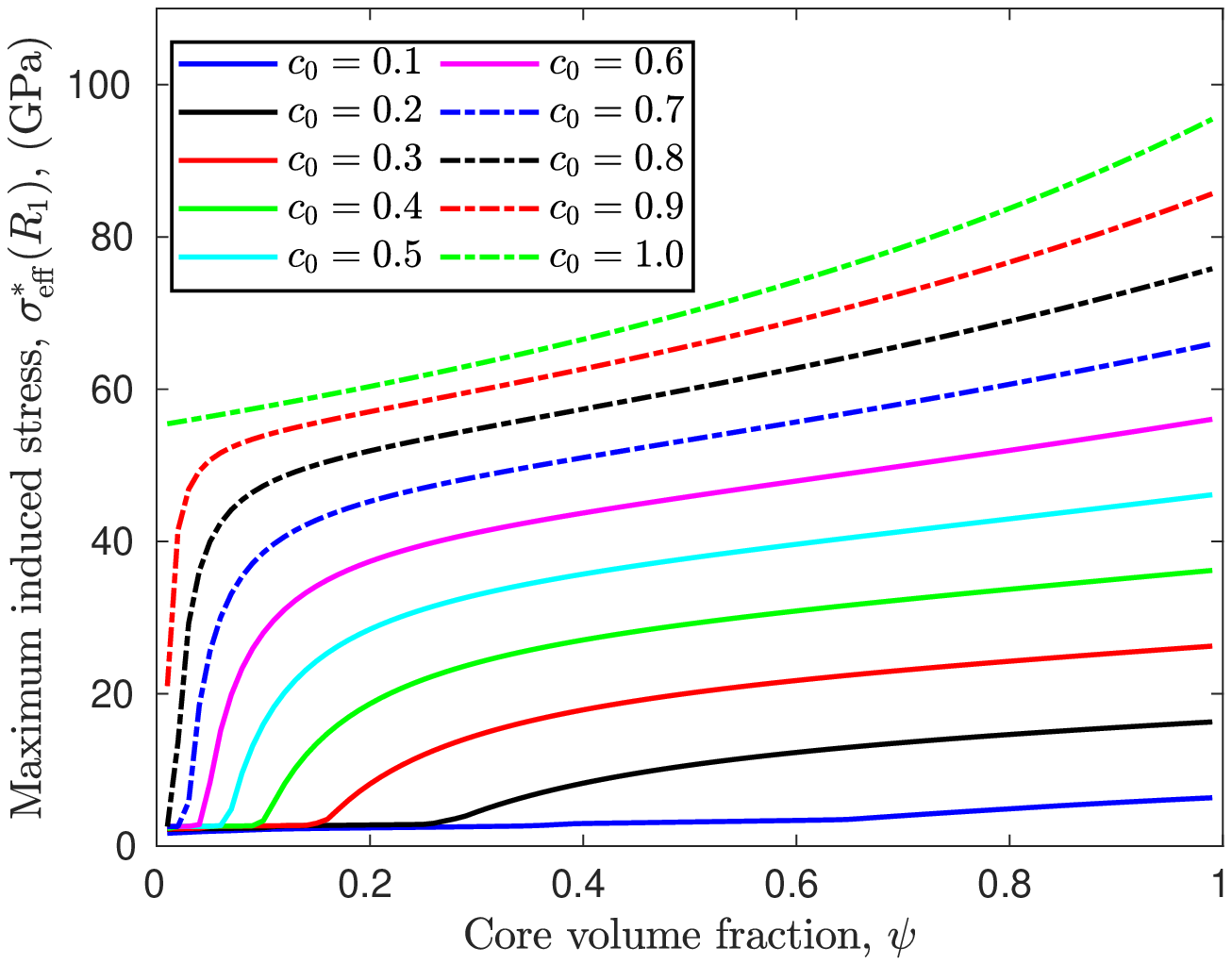}
 \caption{Maximum induced stress, $\sigma_\text{eff}^*(R_1)$, given by \eqref{eq:effective_interfacial_stress}, against core volume fraction, $\volfrac$ for different states of charge, $c_0$.}
 \label{fig:spherical_linear_sigma_eff}
 \end{figure}

 In Figure~\ref{fig:spherical_linear_sigma_eff}, we plot $\sigma_\text{eff}^*(R_1)$ against $\volfrac$ for different SOC.
 For all states of charge, $\sigma_\text{eff}^*(R_1)$ is minimised as $\volfrac \rightarrow 0$; therefore, a very small silicon core is optimal.
 Of course, a single-material anode nano-particle will not induce any stress in equilibrium and thus the two single-material designs in this case are globally optimal for this measure with $\sigma_\text{eff}^*(R_1) = 0$.
 
 \subsection{Optimisation}\label{subsec:optimisation}
 
 We now suggest objective functions and constraints that could be used to optimise the design of an anode using the three performance measures defined above.
 We derive these objective functions and constraints for the core--shell spherical nano-particle shown in Figure~\ref{fig:spherical_geometry}, and plot how the objective function varies with the volume of the silicon core with the aim of finding the optimal size of silicon core for each optimisation problem.

 \subsubsection{Amount of Lithium per Relative Expanded Volume}\label{subsubsec:Amount_of_lithium_per_volume}
 
 The performance of anode materials are often measured by their volumetric or gravimetric capacity.
 The first measure we use to optimise the anode geometry is the amount of lithium per expanded volume which is equivalent to the volumetric capacity.
 We calculate the amount of lithium per expanded volume by dividing the relative amount of lithium, $Q$ in \eqref{eq:Q}, by the relative expanded volume, $\expvol$ in \eqref{eq:spherical_V/V_0}, giving
 \begin{equation}
 \frac{Q}{V} = \frac{c_1 \rad^3 + c_2\frac{c_2^\text{max}}{c_1^\text{max}}(1-\rad^3)}{(1+\eta_1V_1^mc_1^\text{max}u(1))^3}.\label{eq:spherical_QV0_V}
 \end{equation}
 
 \begin{figure}[b!]
 \centering
 \includegraphics[scale=0.79]{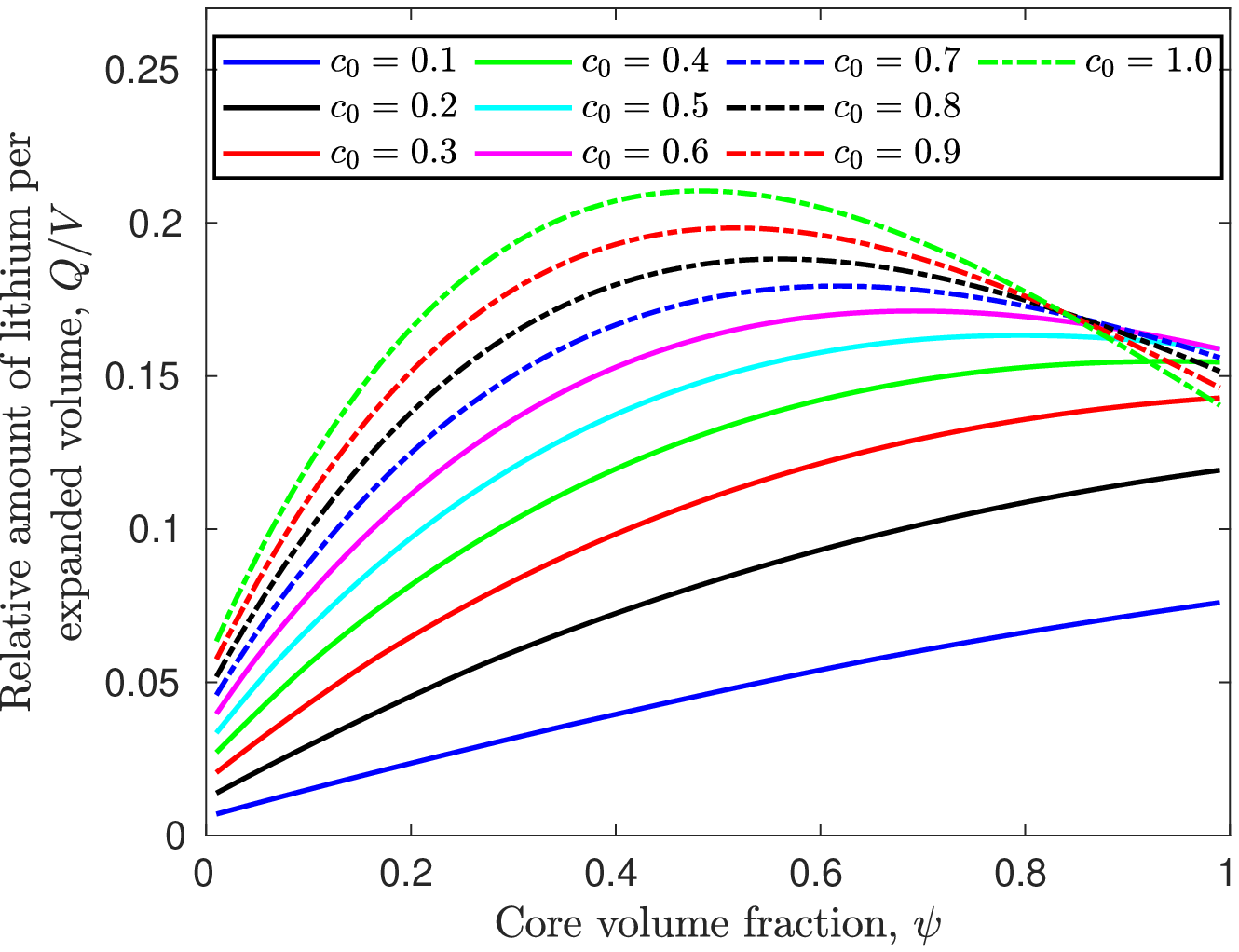}
 \caption{Relative amount of lithium per relative expanded volume, $Q/V$, given by \eqref{eq:spherical_QV0_V}, against core volume fraction, $\volfrac$, for different states of charge, $c_0$.}
 \label{fig:spherical_linear_QV0_V}
 \end{figure}
 
 In Figure~\ref{fig:spherical_linear_QV0_V}, we plot $Q/V$ against $\volfrac$ for the same states of charge as in Figures~\ref{fig:spherical_linear_V/V_0}-\ref{fig:spherical_linear_sigma_eff}.
 We can see from this plot that $Q/V$ is maximised by a fully lithiated nano-particle with a silicon core of volume fraction $\approx 0.45$.
 The reason that $Q/V$ is not optimised by a nano-particle of a single material is that the amount of lithium $Q$ is linear in $\volfrac$ ($= \rad^3$), whereas the expanded volume $\expvol$ increases much more rapidly with $\volfrac$ at large values of $\volfrac$ than for smaller values.
 Therefore, as $\volfrac$ becomes larger, the extra lithium that can be intercalated is out-weighed by the increased expansion and $Q/V$ begins to decrease with $\volfrac$.
 However, as explained in Section~\ref{subsubsec:Relative_Expanded_Volume}, the linear elasticity model overestimates $\expvol$ at high SOC and high $\volfrac$ so this prediction needs to be validated with a nonlinear model.
 For small SOC values, $Q/V$ has a much more linear relationship with $\volfrac$, and thus if the nano-particle is only lithiated a small amount, a fully silicon nano-particle is the optimal design according to the $Q/V$ measure.
 Lastly, we observe that for large $\volfrac$, $Q/V$ is not monotonic with $c_0$ and so having a partially lithiated nano-particle yields a higher volumetric capacity than a fully lithiated one.
 
 \subsubsection{Maximising Amount of Lithium Constrained by Maximum Expanded Volume}\label{subsubsec:expanded_volume_constraint}
 
 We now consider the problem of maximising the amount of lithium the anode can hold subject to constraints on the expanded volume and the maximum effective stress.
 We begin with the expanded volume constraint.
 Thus, we want to maximise $Q$, given by \eqref{eq:general_Q}, subject to $\expvol\leq\expvol_\text{max}$, for some prescribed $\expvol_\text{max}$.
 
 For a given $\volfrac$, $Q$ is maximised by maximising the SOC, $c_0$.
 From Figure~\ref{fig:spherical_linear_V/V_0}, we see that for a given $\volfrac$ value, the expansion $\expvol$ is monotonically increasing with $c_0$.
 Thus, the maximum viable $c_0$ value, which we denote as $\maxconc$ occurs when $\expvol = \expvol_\text{max}$, unless the fully lithiated volume is less than the constraint, in which case $\maxconc = 1$.
 Therefore, for each $\volfrac \in [0,1]$, we find $\maxconc$ such that
 \begin{equation}
 \expvol_{c_0 = \maxconc} = \min\big[\expvol_{c_0=1},\expvol_\text{max}\big].\label{eq:general_V_V0_calc}
 \end{equation}
 We then maximise the corresponding $Q_\text{max} = Q(\maxconc)$ over $\volfrac$.

 \begin{figure}[b!]
 \centering
 \includegraphics[scale=0.79]{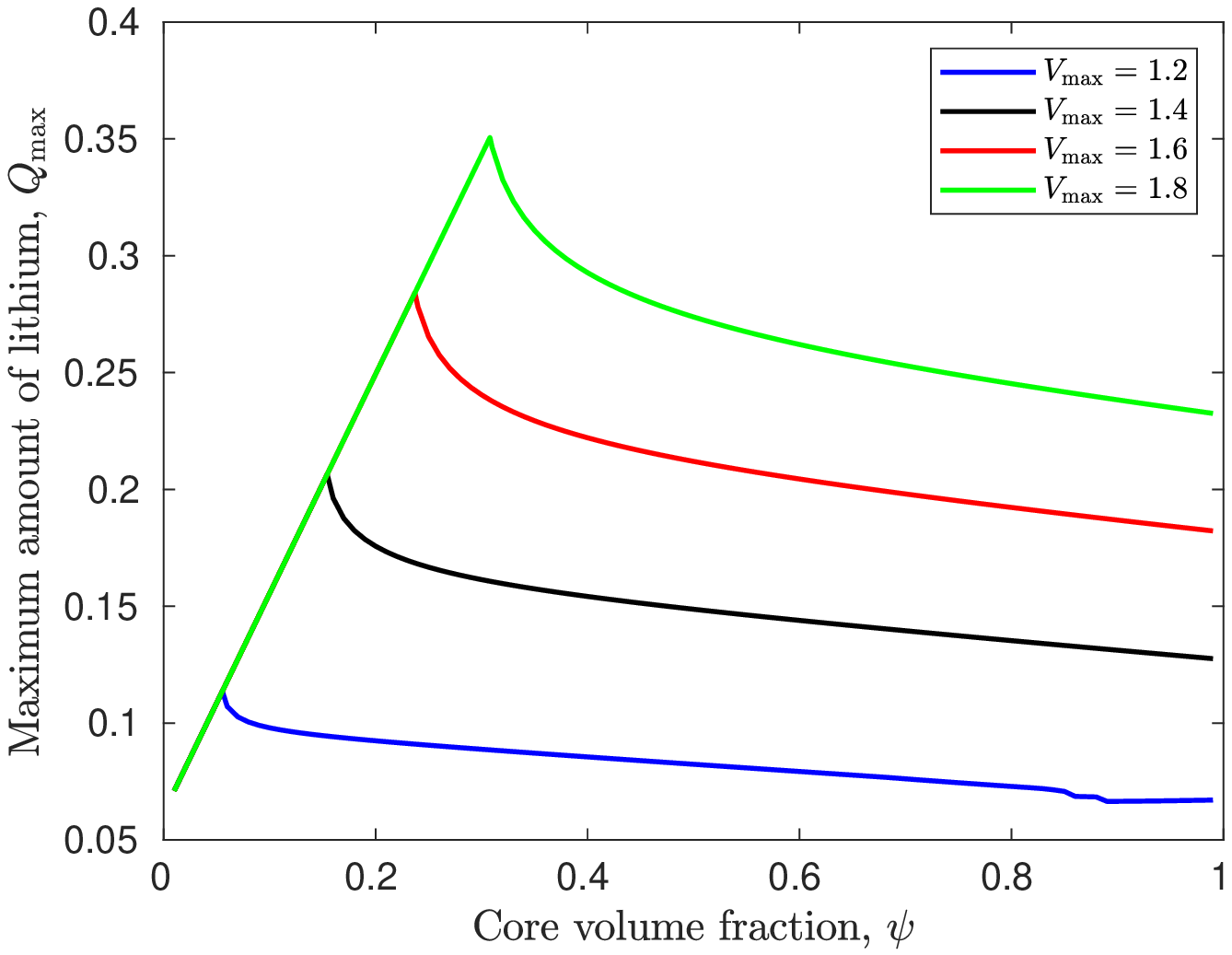}
 \caption{Maximum amount of lithium $Q_\text{max}$ against core volume fraction $\volfrac$ for different values of the maximum permitted expanded volume $(V/V_0)_\text{max}$.}
 \label{fig:spherical_linear_Q_for_max_expansion}
 \end{figure}
 \begin{figure}[t!]
 \centering
 \includegraphics[scale=0.79]{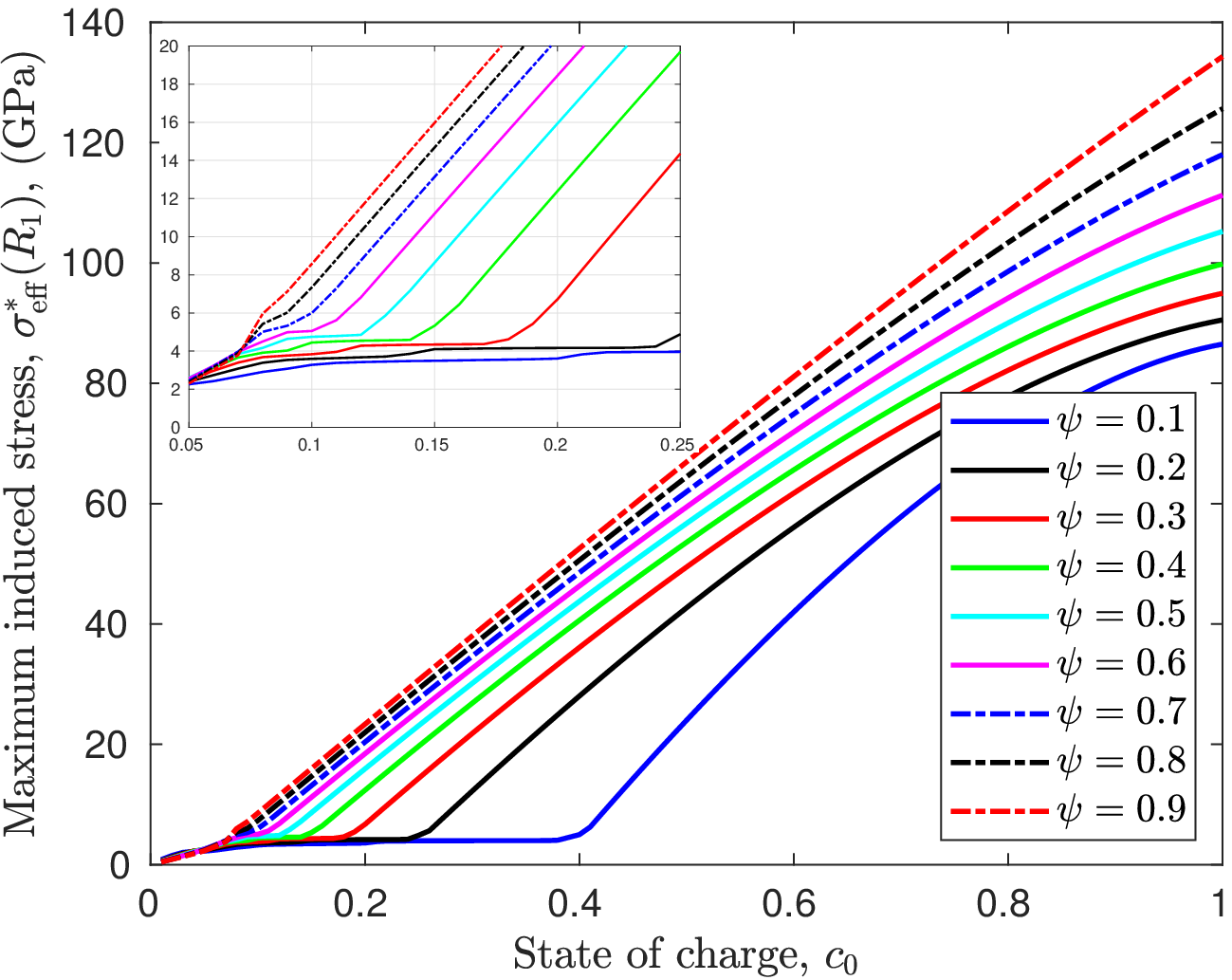}
 \caption{Maximum induced stress, $\sigma_\text{eff}^*(R_1)$, against state of charge, $c_0$, for different core volume fractions $\volfrac$. The inset plot is focused on $0<\sigma_\text{eff}^*(R_1)<20.0$ GPa to show more clearly that it is monotonically increasing in $c_0$.}
 \label{fig:eff_stress_vs_c0}
 \end{figure}
 
 The value of $\volfrac$ at which $\expvol_{c_0=\maxconc}$ changes from $\expvol_{c_0=1}$ to $\expvol_\text{max}$ in \eqref{eq:general_V_V0_calc} can be calculated by substituting \eqref{eq:linear_elasticity_solution} into \eqref{eq:spherical_V/V_0} for $c_1 = c_2 = 1$ and equating to the prescribed $\expvol_\text{max}$, giving
 \begin{equation}
 \big[1 + \eta_1V_1^mc_1^\text{max}\big(\hat{A}_2 + \hat{B}_2\big)\big]^3 = \expvol_\text{max},\label{eq:V/V_0_equation_in_A_and_B_max}
 \end{equation}
 where $\hat{A}_2$ and $\hat{B}_2$ are given by \eqref{eq:linear_spherical_A2} and \eqref{eq:linear_spherical_B2} with $c_1 = c_2 = 1$.
 This can be solved to find a critical value of $\rad$, which we denote $\hat{\rad}$, which corresponds to the $\volfrac$ value at which $\expvol_{c_0=1}=\expvol_\text{max}$.
 For all values of $\volfrac$ such that $\rad<\hat{\rad}$, $Q_\text{max}$ is given by \eqref{eq:Q} with $c_1 = c_2 = 1$ and $\rad = \volfrac^{1/3}$.
 However, for $\rad>\hat{\rad}$, we must numerically solve
 \begin{equation}
 \big[1 + \eta_1V_1^mc_1^\text{max}(A_2 + B_2)\big]^3 = \bigg(\frac{V}{V_0}\bigg)_\text{max},\label{eq:V/V_0_equation_in_A_and_B}
 \end{equation}
 to find $\maxconc$ and $Q_\text{max}$.
 We both derive the expression for $\hat{\rad}$ and the bounds on $V_\text{max}$ such that $0 < \hat{R} < 1$ in \ref{app:max_lithium_for_max_expansion}.
  
 In Figure~\ref{fig:spherical_linear_Q_for_max_expansion} we plot $Q_\text{max}$ against $\volfrac$ for four different values of $\expvol_\text{max}$.
 The $\volfrac$ values for which $\rad<\hat{\rad}$ can be seen by the straight line on the left side of the plot.
 This is because for $c_1 = c_2 = 1$, $Q$ is a linear function of $\rad^3 = \volfrac$ in \eqref{eq:Q}.
 Immediately to the right of this region, $Q_\text{max}$ decreases for $R > \hat{\rad}$.
 The $\volfrac$ value which gives the largest $Q_\text{max}$ is $\volfrac = \hat{\rad}^3$ as can be seen by the peaks in $Q_\text{max}$ after the linear region for small $\volfrac$.
 
 \subsubsection{Maximising Amount of Lithium Constrained by Maximum Stress}
 
 We now wish to maximise the amount of intercalated lithium under a constraint on the maximum induced stress, defined in \eqref{eq:general_von_mises}.
 We wish to find the anode geometry which maximises \eqref{eq:general_Q} subject to the maximum induced stress in the anode \eqref{eq:effective_interfacial_stress} being less than some prescribed maximum $\sigma_\text{max}$.
 
 As with the expanded volume constraint in Section~\ref{subsubsec:expanded_volume_constraint}, for each volume fraction of silicon, $\volfrac$, we must find the maximum $c_0$ such that the maximum induced stress $\sigma_\text{eff}^*(R_1)$ is less than $\sigma_\text{max}$.
 In Figure~\ref{fig:eff_stress_vs_c0}, we plot $\sigma_\text{eff}^*(R_1)$ against SOC and it can be seen that $\sigma_\text{eff}^*(R_1)$ is a monotonically increasing function of $c_0$.
 Therefore, the maximal $c_0$ value, denoted by $\maxconc$, under the constraint $\sigma_\text{eff}^*(R_1) < \sigma_\text{max}$ will be such that $\sigma_\text{eff}^*(R_1) = \sigma_\text{max}$.
 We then calculate the maximum amount of lithium, $Q_\text{max} = Q(\maxconc)$, for that $\volfrac$ value by substituting $\maxconc$ into \eqref{eq:Q}.
 
 \begin{figure}[t!]
 \centering
 \includegraphics[scale=0.79]{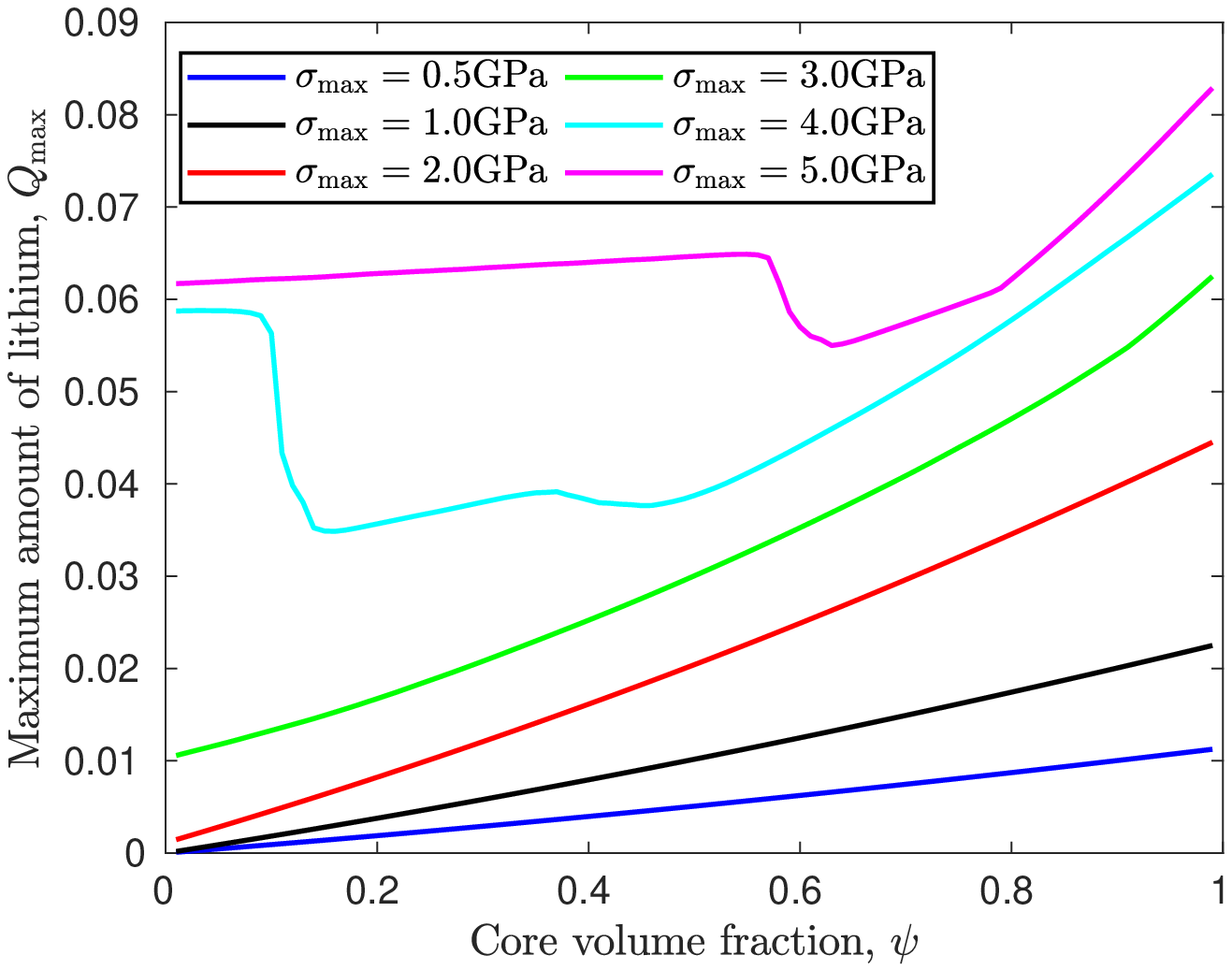}
 \caption{Maximum amount of lithium, $Q_\text{max}$, against core volume fraction $\volfrac$ for different values of the maximum permitted effective stress $\sigma_\text{max}$.}
 \label{fig:spherical_linear_Q_for_max_effective_stress}
 \end{figure} 
 
 In Figure~\ref{fig:spherical_linear_Q_for_max_effective_stress}, we plot $Q_\text{max}$ against $\volfrac$ for several different values of $\sigma_\text{max}$.
 It can be seen that $\volfrac = 1$ gives the greatest $Q_\text{max}$ value for each value of $\sigma_\text{max}$.
 However, for $\sigma_\text{max} =4.0$ and 5.0 GPa, $Q_\text{max}$ is not monotonic in $\volfrac$, thus there are local maxima of $Q_\text{max}$.
 This is useful for design purposes given other constraints. 
 For example if the maximum stress that is allowed is $\sigma_\text{max} = 4.0$ GPa, but we must also restrict $\volfrac$ to be less than 0.25, having $\volfrac = 0.05$ would allow a larger amount of lithium to be intercalated than with $\volfrac = 0.2$.
 
 As in Section~\ref{subsubsec:expanded_volume_constraint}, for sufficiently large $\sigma_\text{max}$, there are $\volfrac$ values for which a fully lithiated nano-particle ($\maxconc = 1$) does not produce stresses as large as $\sigma_\text{max}$.
 Therefore, the the linear relationship for small $\volfrac$ in Figure~\ref{fig:spherical_linear_Q_for_max_expansion} is also seen for large values of $\sigma_\text{max}$.
 However, the $\sigma_\text{max}$ values illustrated in Figure~\ref{fig:spherical_linear_Q_for_max_effective_stress} are too small to see
this behaviour; all of the curves there have $\maxconc < 1$.
 We give the value of $\hat{R}$ (the core radius for which the maximum induced stress at $c_0=1$ is $\sigma_\text{max}$) and the bounds on $\sigma_\text{max}$ such that $0<\hat{R}<1$ (thus the linear relationship for small $\volfrac$ is observed) in \ref{app:max_lithium_for_max_expansion}.
  
 The $\sigma_\text{max}$ values plotted for here are much greater than the tensile strength of graphite, $\sigma_\C^Y$, typically between 8 and 12 MPa \cite{Manhani2007}.
 The results for the constraint $\sigma_\text{max} = \sigma_\C^Y$ are qualitatively similar to that of $\sigma_\text{max} = 0.5$ GPa.
 Therefore, to avoid the yielding or cracking of the graphite at the interface, the optimal design is as large a silicon core as possible.
 For $\volfrac = 0.99$, the silicon can only be lithiated to around $c_1 = 2.2\times 10^{-4}$ for the interfacial stress \eqref{eq:effective_interfacial_stress} to remain under the yield stress, while the graphite remains unlithiated.
 This achieves a $Q_\text{max}$ value also of around $2.2\times 10^{-4}$ (as $c_2 = 0$ in this case).

 \section{Conclusions and Discussion}\label{sec:conclusions}
 
 In this paper, we present a model for the inclusion of mechanical stress in calculating lithium distribution for multi-material lithium-ion batteries.
 We show how the stress can be included into the concentration model for simple geometries; for example spheres, cylinders and 1-dimensional plates, and determine that the lithium concentration is uniform in each material when diffusion is fast compared to the charging rate.
 By applying the model to a radially symmetric spherical nano-particle with a silicon core and a graphite shell, we show there are large changes in lithium distribution if the chemo-mechanical coupling effects are neglected.
 Finally, we present a framework for finding the optimal design for the geometry of multi-material anodes and present three performance measures and three optimality conditions.
 We present results from optimising the core volume for a silicon core, graphite shell geometry as an example of the insight that can be obtained from using these measures.
 
 There are several limitations to the static linear elasticity model we derive in Section~\ref{sec:model}.
 Firstly, while the quasistatic assumption is valid for small currents, the non-uniformity in the lithium concentration caused by diffusion through the anode during more strenuous battery function causes substantial stresses that are not captured by this model.
 Secondly, there are mechanical phenomena commonly observed experimentally, such as plasticity and cracking, that are also not included in this model.
 Our solely elastic mechanical model assumes that the materials are able to act elastically regardless of the stress induced by the lithiation and there is no yield stress at which the material either begins to plasticise or crack.
  Lastly, as noted in Section~\ref{subsec:nondimensionalisation}, the linear elasticity assumptions used to derive the mechanical model in Section~\ref{subsubsec:Mechanical_model} relies on the parameter $\eta_a V_a^m c_a^\text{max} \ll 1$ for $a = 1,\dots,n$, thus restricting the materials this linear model is applicable to.
 
 We also make several assumptions about the materials in our model which are significant simplifications for many anode materials.
 We assume the anode materials are isotropic to allow us to write the stiffness tensor in terms of two Lam\'{e} parameters, $\lambda$ and $G$, and this ignores the anisotropy of the crystal structure of the anode.
 While we have taken the lithium-concentration-dependence of the Young's modulus into account, we still assume that the Poisson's ratio and molar density are lithium-concentration-independent.
 Finally, we use parameters and a model suitable for bulk materials, while we are imagining nano-structures in our example.
 It has been experimentally shown that nano-sized materials have very different mechanical properties from bulk materials \cite{zhu2009mechanical} and accounting for surface effects in nano-structures becomes increasingly important as the size of the material decreases \cite{roduner2006size}.
 
 We use the example of a silicon core and a graphite shell to show the results in Figures~\ref{fig:spherical_linear_conc_vs_c0}-\ref{fig:spherical_linear_Q_for_max_effective_stress}; however, for silicon, $\eta_1 V_1^m c_1^\text{max} = 0.933375$.
 Therefore, our application of this model to a silicon core surrounded by a graphite shell needs to be interpreted with some care.
 The expanded volume of the hybrid particle with a silicon core and graphite shell, shown in Figure~\ref{fig:spherical_linear_V/V_0}, is larger than the expected values because of the linear assumption being violated.
 Additionally, the disparity between $\eta_1$ and $\eta_2$ in Table~\ref{table:CS_parameters} means that the graphite shell is being stretched by a large amount, causing very large tensile stresses.
 As we have neglected cracking and plasticity from the model, this means the tensile stresses of the graphite are most likely exaggerated in the results shown here, which will not only affect the optimisation results in Section~\ref{sec:optimal_size} but also the lithium distributions within the anode.
 Although the model is somewhat outside its range of validity when applied to silicon anodes, we use silicon as material 1 in our example to clearly show the necessity of including stress into the static lithium concentration model due to its large expansion.
 This model can still be applicable to silicon if we were to restrict the lithiation of the silicon such that $V^m_1 c_1^\text{max}<3.75$, thus the results shown here are still accurate for small $c_1$ levels.
 
 Despite the limitations of the model, the simplifications made in this work allow us to analytically show the consequences of including a stress-dependence into the chemical potential.
 One key result of this is to show that stress is induced solely by the presence of different materials, independently of the non-uniformity of the lithium concentration due to diffusion.
 While the differences in stress-dependent lithium concentrations and stress-independent concentrations will be less pronounced with materials with lower expansions than silicon, these stresses are still important to include for accurate OCV prediction and stress modelling.
 While several anode materials have non-isotropic crystal structures, causing the isotropy assumption to be inappropriate, the crystalline structure of anode materials usually become amorphous after the first charge, especially if the stresses are large.
 Additionally, the introduction of $\lambda$ and $G$ does not hugely simplify the model for general geometries and so removing this assumption for these cases would only slightly increase the complexity of the model.
 Finally, the simplifications made to the model have allowed to provide analytical insight into the key performance indicators of anode nano-particles that we define in Section~\ref{sec:optimal_size}.
 Therefore, this model can be used for insight into optimal anode design for lithium-ion batteries, but it must be adapted for more quantitatively accurate results and for more challenging design geometries to be used in practice.
 
\section{Acknowledgements}
 
 This publication is based on work supported by the EPSRC Centre For Doctoral Training in Industrially Focused Mathematical Modelling (EP/L015803/1) in collaboration with Nexeon.
 Special acknowledgement goes to the industrial collaborator, Bill Macklin, for very insightful discussions and motivations for this work. JC and CP acknowledge funding
from the Faraday Institution (EP/S003053/1).
 
 \appendix
 \section{Calculation of $\hat{\rad}$ for Maximum Expansion and Maximum Stress Constraints}\label{app:max_lithium_for_max_expansion}
 
 We present expressions for the critical radii $\hat{\rad}$ for the maximum expansion constraint and the maximum stress constraint where $\hat{\rad}$ is the radius of the silicon core such that the constraint is met when $c_0 = 1$.
 We also present the inequalities that the constraints $V_\text{max}$ and $\sigma_\text{max}$ must satisfy such that $0<\hat{\rad}<1$.
 
 \subsection{Maximum Expansion Constraint}
 
 Substituting \eqref{eq:linear_spherical_A2}-\eqref{eq:linear_spherical_B2} into \eqref{eq:V/V_0_equation_in_A_and_B_max} yields
 \begin{align}
  \frac{\bar{\eta}_1\big[\Lambda_1\Lambda_2\big(\hat{\rad}^3 \gamma_1 + (1-\hat{\rad}^3)\gamma_2\big) + 4G_2 \big(\Lambda_1\hat{\rad}^3\gamma_1 + \Lambda_2(1-\hat{\rad}^3)\gamma_2\big)\big]}{\Lambda_1\Lambda_2+ 4G_2\big(\Lambda_2(1-\hat{\rad}^3) + \Lambda_1\hat{\rad}^3\big)} =\expvol^\frac{1}{3}_\text{max} - 1,
 \end{align}
 where $\Lambda_a = 3\lambda_a + 2G_a$ for $a = 1,2$ and $\bar{\eta}_1 = \eta_1 V_1^m c_1^\text{max}$.
 This can be rearranged to give
 \begin{equation}
 \hat{\rad} = \Bigg[\frac{(\Lambda_1\Lambda_2 + 4G_2\Lambda_2)\Big(V_\text{max}^{1/3} - 1\Big) - \bar{\eta}_1\Lambda_2\gamma_2(\Lambda_1 + 4G_2)}{\bar{\eta}_1\big[\Lambda_1\Lambda_2(\gamma_1-\gamma_2) + 4G_2(\Lambda_1\gamma_1 - \Lambda_2\gamma_2)\big] - 4G_2(\Lambda_1-\Lambda_2)\Big(V_\text{max}^{1/3} - 1\Big)}\Bigg]^\frac{1}{3}.\label{eq:critical_rad}
 \end{equation}
 We rearrange the inequality $0<\hat{\rad}<1$ with $\hat{\rad}$ given by \eqref{eq:critical_rad} to find
 \begin{equation}
 \bigg(\frac{\bar{\eta}_1 \Lambda_2 \gamma_2(\Lambda_1+4G_2)}{\Lambda_1\Lambda_2+4G_2\Lambda_2} + 1\bigg)^3 < V_\text{max} < (1+\bar{\eta}_1\gamma_1)^3.
 \end{equation}
 
 \subsection{Maximum Stress Constraint}
 
 Substituting \eqref{eq:linear_spherical_B2} into \eqref{eq:effective_interfacial_stress} with $c_1 =c_2 = c_0 = 1$ yields
 \begin{equation}
 \sigma_\text{max} = \frac{6\eta_1V_1^mc_1^\text{max}G_2^* \Lambda_1 \Lambda_2 |\gamma_1 - \gamma_2|}{\Lambda_1 \Lambda_2 + 4G_2\big(\Lambda_2(1-\hat{\rad}^3) + \Lambda_1 \hat{\rad}^3\big)},
 \end{equation}
 which can be rearranged to give
 \begin{equation}
 \hat{\rad} = \Bigg[\frac{\Lambda_1\Lambda_2(6\eta_1V_1^mc_1^\text{max}G_2^*|\gamma_1 - \gamma_2| - \sigma_\text{max}) - 4G_2\Lambda_2\sigma_\text{max}}{4G_2(\Lambda_1-\Lambda_2)\sigma_\text{max}}\Bigg]^\frac{1}{3}.\label{eq:stress_R_hat}
 \end{equation}
 We rearrange $0<\hat{\rad}<1$ with $\hat{\rad}$ now given by \eqref{eq:stress_R_hat} to find
 \begin{equation}
 \frac{6\eta_1V_1^mc_1^\text{max}G_2^*\Lambda_1\Lambda_2|\gamma_1-\gamma_2|}{\Lambda_1\Lambda_2 + 4G_2\Lambda_2} < \sigma_\text{max} < \frac{6\eta_1V_1^mc_1^\text{max}G_2^*\Lambda_1\Lambda_2|\gamma_1 - \gamma_2| - 4G_2(\Lambda_1-\Lambda_2)}{\Lambda_1\Lambda_2 + 4G_2\Lambda_2},
 \end{equation}
 must be satisfied to satisfy $0<\hat{\rad}<1$.

%
%



\end{document}